\documentclass[12pt]{article}
\usepackage{graphicx}
\usepackage{amsmath}

\newcommand{\nc}{\newcommand}
\nc{\bq}{\begin{equation}}
\nc{\eq}{\end{equation}}
\nc{\bqy}{\begin{eqnarray}}
\nc{\eqy}{\end{eqnarray}}
\nc{\p}{\partial}

\nc{\om}{\omega}
\nc{\Om}{\Omega}
\nc{\de}{\delta}
\nc{\al}{\alpha}
\nc{\be}{\beta}
\nc{\ga}{\gamma}
\nc{\ka}{\kappa}
\nc{\ze}{\zeta}
\nc{\si}{\sigma}
\nc{\up}{\Upsilon}
\nc{\la}{\lambda}

\nc{\omp}{\ze}

\nc{\psek}{{\psi_{e}}_{k}}
\nc{\psemk}{{\psi_{e}}_{-k}}

\nc{\ptk}{{T_{+}}_{k}}
\nc{\ptmk}{{T_{+}}_{-k}}
\nc{\mtk}{{T_{-}}_{k}}
\nc{\mtmk}{{T_{-}}_{-k}}

\nc{\ptl}{{T_{+}}_{\ell}}
\nc{\ptml}{{T_{+}}_{-\ell}}
\nc{\mtl}{{T_{-}}_{\ell}}
\nc{\mtml}{{T_{-}}_{-\ell}}

\nc{\fw}{\frac{\delta F}{\delta \tilde \omp}}
\nc{\gw}{\frac{\delta G}{\delta \tilde \omp}}

\nc{\fd}{\frac{\delta F}{\delta \tilde D}}
\nc{\gd}{\frac{\delta G}{\delta \tilde D}}

\nc{\fp}{\frac{\delta F}{\delta \tilde T_+}}
\nc{\gp}{\frac{\delta G}{\delta \tilde T_+}}
\nc{\fpm}{\frac{\delta F}{\delta \tilde T_\pm}}
\nc{\gpm}{\frac{\delta G}{\delta \tilde T_\pm}}

\nc{\fm}{\frac{\delta F}{\delta \tilde T_-}}
\nc{\gm}{\frac{\delta G}{\delta \tilde T_-}}

\nc{\lps}{L_{\psi}\bar\psi_0'}
\nc{\lph}{L_{\phi}\bar\phi_0'}
\nc{\lv}{L_{v}\bar{v}_0'}
\nc{\lz}{L_{Z}\bar{Z}_0' }
\nc{\ld}{L_{D}\bar{D}_0' }
\nc{\lom}{L_{\omp}\bar{\omp}_0' }
\nc{\ltp}{L_{T_+}\bar{T}_{+0}' }
\nc{\ltm}{L_{T_-}\bar{T}_{-0}' }
\nc{\ltpm}{L_{T_{\pm}}\bar{T}_{\pm 0}' }

\nc{\py}{\frac{\p }{\p y}}

\nc{\st}{\sum_{k=-\infty}^{\infty}}
\nc{\su}{\sum_{k=1}^{\infty}}

\nc{\intp}{\int_{-\pi}^{\pi}\!d\,y}

\nc{\dff}{\frac{\delta F}{\delta   f}}
\nc{\dgf}{\frac{\delta G}{\delta   f}}

\nc{\dbfk} {\frac{\p \bar{F}}{\p f_k}}
\nc{\dbfmk} {\frac{\p \bar{F}}{\p f_{-k}}}
\nc{\dfk} {\frac{\p F}{\p f_k}}
\nc{\dfmk} {\frac{\p F}{\p f_{-k}}}

\nc{\fdk}{\frac{\p \bar F}{\p  D_{k}}}
\nc{\fdmk}{\frac{\p \bar F}{\p  D_{-k}}}
\nc{\gdk}{\frac{\p \bar G}{\p  D_{k}}}
\nc{\gdmk}{\frac{\p \bar G}{\p  D_{-k}}}

\nc{\fok}{\frac{\p \bar F}{\p  \omp_{k}}}
\nc{\fomk}{\frac{\p \bar F}{\p  \omp_{-k}}}
\nc{\gok}{\frac{\p \bar G}{\p  \omp_{k}}}
\nc{\gomk}{\frac{\p \bar G}{\p \omp_{-k}}}

\nc{\fptk}{\frac{\p \bar F}{\p  \ptk}}
\nc{\fptmk}{\frac{\p \bar F}{\p  \ptmk}}
\nc{\gptk}{\frac{\p \bar G}{\p  \ptk}}
\nc{\gptmk}{\frac{\p \bar G}{\p \ptmk}}

\nc{\fmtk}{\frac{\p \bar F}{\p  \mtk}}
\nc{\fmtmk}{\frac{\p \bar F}{\p  \mtmk}}
\nc{\gmtk}{\frac{\p \bar G}{\p  \mtk}}
\nc{\gmtmk}{\frac{\p \bar G}{\p \mtmk}}

\nc{\fdl}{\frac{\p \bar F}{\p  D_{\ell}}}
\nc{\fdml}{\frac{\p \bar F}{\p  D_{-\ell}}}
\nc{\gdl}{\frac{\p \bar G}{\p  D_{\ell}}}
\nc{\gdml}{\frac{\p \bar G}{\p  D_{-\ell}}}

\nc{\fol}{\frac{\p \bar F}{\p  \omp_{\ell}}}
\nc{\foml}{\frac{\p \bar F}{\p  \omp_{-\ell}}}
\nc{\gol}{\frac{\p \bar G}{\p  \omp_{\ell}}}
\nc{\goml}{\frac{\p \bar G}{\p \omp_{-\ell}}}

\nc{\fptl}{\frac{\p \bar F}{\p  \ptl}}
\nc{\fptml}{\frac{\p \bar F}{\p  \ptml}}
\nc{\gptl}{\frac{\p \bar G}{\p  \ptl}}
\nc{\gptml}{\frac{\p \bar G}{\p T_{+-\ell}}}

\nc{\fmtl}{\frac{\p \bar F}{\p  \mtl}}
\nc{\fmtml}{\frac{\p \bar F}{\p  \mtml}}
\nc{\gmtl}{\frac{\p \bar G}{\p  \mtl}}
\nc{\gmtml}{\frac{\p \bar G}{\p \mtml}}

\nc{\cb}{c_{\beta}}
\nc{\db}{d_{\beta}}

\nc{\als}{\al_{\psi}}
\nc{\ald}{\al_D}
\nc{\alp}{\al_+}
\nc{\alm}{\al_{-}}
\nc{\alo}{\al_{\omp}}
\nc{\alv}{\al_v}
\nc{\alz}{\al_Z}

\nc{\albd}{\alpha_{\bar D}}
\nc{\aalbd}{|\alpha_{\bar D}|}

\nc{\kat}{\ka_{\perp}}
\nc{\kt}{k_{\perp}}

\begin{document}
\title{Hamiltonian formulation  and analysis of a collisionless fluid 
reconnection model}
\author{E. Tassi$^*$ \\
Dipartimento di Energetica, Politecnico di Torino,\\
Torino, 10129, Italy\\
$^*$E-mail: emanuele.tassi@polito.it\\
\\
P. J. Morrison and F.\ L.\ Waelbroeck\\
Department of Physics and Institute for Fusion Studies, University of Texas\\
Austin, Texas 78712, United States\\
\\
D. Grasso\\
Dipartimento di Energetica, Politecnico di Torino,\\
Torino, 10129, Italy\\}

\maketitle

\begin{abstract}
The Hamiltonian formulation of a plasma  four-field fluid model  that describes 
collisionless reconnection is presented. The formulation  is noncanonical with a corresponding Lie-Poisson bracket.  The bracket  is used  to  obtain new independent families of invariants, so-called Casimir invariants, three  of which are directly related to Lagrangian invariants of the system.  The Casimirs are used to obtain a variational principle for equilibrium equations that generalize the Grad-Shafranov equation to  include flow.   Dipole and homogeneous equilibria are constructed.  The linear dynamics of the latter is treated in detail in a Hamiltonian context:  canonically conjugate variables are obtained;  the dispersion relation is analyzed and exact thresholds for spectral stability are obtained;  the canonical transformation to normal form is described;  an unambiguous definition of negative energy modes is given;  and  thresholds  sufficient  for energy-Casimir stability are obtained. The Hamiltonian  formulation also is used to obtain an  expression for the collisionless conductivity and it is further used to  describe the linear growth and nonlinear saturation of the collisionless tearing mode.  
\end{abstract}


\maketitle

\section{Introduction} \label{sec:intro}

Due to the lower dimensionality of configuration space as compared to 
phase space, fluid models of the plasma have an intrinsic 
computational advantage over kinetic models. For this reason it is of 
great interest to develop and apply such models even in cases where 
the collisionality is too small to provide a firm justification for 
their use. In particular, fluid models have made important 
contributions to the understanding of magnetic reconnection, a phenomenon  
that plays a key role in events such as solar flares, magnetospheric substorms, and sawtooth 
oscillations in tokamaks. \cite{RogrDentDrak,Pri00,Bis00,Sch94} They have also made key contributions to the understanding of plasma turbulence in the core  \cite{DastMaWeilnd,KromKolesh,KolesKromPRL,KolesKromPoP} and 
edge\cite{Xu_L-H,RogrDrakZeilr,GuzMaYo,ScottRvw08} of magnetic confinement experiments 
in collisionless as well as collisional regimes. More generally, they 
offer the promise of being able to perform simulations of multiscale 
phenomena that are beyond the reach of kinetic models even after 
accounting for foreseeable advances in computation speed.\cite{ScottGEM06}

An important consideration when constructing new plasma fluid models is the 
existence of a Hamiltonian structure (see  \cite{mor82,mor98,mor05} for reviews).
Because the fundamental  laws governing charged particle dynamics are Hamiltonian, dissipative terms, which ultimately arise from simplifications,  must be accompanied by phenomenological constants  such as resistivity and viscosity.  When such phenomenological quantities are neglected, it is desirable that the resulting model be Hamiltonian, as is the case for the most important kinetic and fluid models of plasma physics.  The preservation of the Hamiltonian structure provides some confidence that the truncations that are used to derive the fluid model have not introduced unphysical sources of dissipation. The presence of the Hamiltonian structure has the additional benefit of providing important tools for calculations.  For example, the magnetohydrodynamic (MHD) energy principle is a consequence of the  Hamiltonian nature of MHD. 

Since the discovery of the noncanonical Hamiltonian structure of MHD \cite{MG80}, many plasma fluid models have been shown to possess a Hamiltonian description in terms of noncanonical Poisson brackets, e.g.\  
\cite{Sch94,KromKolesh,KolesKromPRL,KolesKromPoP,MH84,mar_mor84,Haz87,Kuv94,Gra01,Gra99}.   In some cases, the requirement  that the dynamics has Hamiltonian form  has been used to guide  the construction\cite{KromKolesh,Haz87} and has led to the  identification of new and physically important terms.\cite{Haz87}   In  another case, the absence of a Hamiltonian structure for a given  model was shown to lead to the violation of the solubility conditions  for the equilibrium equations.\cite{WaelMorHor} The Hamiltonian  structure of fluid models has also been shown to be important for the  consistent calculation of zonal flow dynamics.\cite{KromKolesh,KolesKromPRL,KolesKromPoP}

Several fluid models have been proposed to study electromagnetic plasma dynamics (see 
\cite{ScottRvw08,HM,Porcelrvw} for reviews).  Some of these models have been instrumental in advancing our understanding of magnetic 
reconnection.\cite{Haz87,Kuv94,HM,KuvLakPegSchep,Sch94} In particular, the model of 
Ref.\cite{Haz87} led to the discovery of fast (compared to Sweet-Parker) magnetic reconnection by Aydemir.\cite{AYA,KDW} This model included the effects of finite ion temperature, but neglected electron inertia. An alternative model that neglected ion temperature but did include electron inertia as well as curvature effects was proposed around the same time by Hazeltine and Meiss (HM).  The HM model was originally used  to provide a unified description of the formation of current channels in semi-collisional and collisionless regimes. Its Hamiltonian nature, however, has not been investigated until now. 

Interest in the effects of electron inertia was recently revived by a controversy over its influence on the rate of collisionless magnetic reconnection (CMR). In CMR, the ``frozen-in'' condition of MHD is broken by the inclusion of electron inertia instead of resistivity.\cite{Por93,Caf98} This led Schep and collaborators to study two models that may be viewed as limiting cases of the HM model (aside from the fact that they avoid the Boussinesq approximation, retaining instead the density $n$ in the form $\log n/n_0$).\cite{Sch94,Kuv94,KuvLakPegSchep} These authors constructed a Hamiltonian formulation for these models, which were subsequently used to  demonstrate the role of phase mixing of the Lagrangian invariants during fast reconnection\cite{Gra01} and to investigate the role of instabilities of the nonlinearly developed current sheet.\cite{DelsCaliPeg} More recently, Fitzpatrick and Porcelli (FP) have considered another limiting form of the HM model that, compared to the model derived by Schep {\em et al.}, is valid for a wider range of values of $\beta$, the latter indicating the ratio between the plasma pressure and the magnetic pressure based on the toroidal guide field. The FP model also extends the models of Schep {\em et al.} by including the effects of parallel ion compressibility. It has subsequently been used to study two-fluid effects on the Taylor problem\cite{VeksBian} and on the linear growth of tearing modes.\cite{BianVeks} Rogers {\em et al.} have recently shown that the predictions of the  FP model for the linear growth rate of the tearing mode are in good agreement with those obtained with the gyrokinetic code GS2 when the ion temperature is not too large.\cite{RogrKobaRic}

 In the present  paper,  we investigate the Hamiltonian  structure of the FP version of the model of HM.\cite{HM,Fit04}  The paper is organized as follows. In Sec.~\ref{ham_form} we present the noncanonical Poisson bracket and show that this bracket produces the equations of motion with the appropriate Hamiltonian.  In Sec.~\ref{cassies}, we use the noncanonical Poisson bracket to obtain four infinite families of new Casimir invariants,  three of which suggest  that specific combinations of the field variables are Lagrangian invariants.  In terms of these variables, the equations of motion  and the Hamiltonian structure  achieve a much simplified form.  A preliminary version of these results was   announced in \cite{T07}. 

The remaining sections describe a variety of applications that rely on the Hamiltonian structure.  In Sec.~\ref{equilibria} we describe a variational principle for equilibria of the system and show that they are governed by a generalized Grad-Shafranov system of a pair of coupled  elliptic  equations.  We treat two examples:  dipole equilibria with Bessel function solutions and homogeneous equilibria that support wave motion.   Sec.~\ref{normal} explores the latter example in further detail by showing how to construct conventional canonical variables for the linear dynamics.  We show that the system possesses Alfv\'{e}n-like and drift-shear modes, obtain exact stability thresholds, and  give a definition of negative energy modes. We note that the Hamiltonian form is indispensable for an unambiguous definition of negative energy modes.   We also obtain energy-stability conditions, sufficient  conditions for stability  akin to the $\de W$ criterion of MHD. Section \ref{tearing} contains a derivation of the collisionless conductivity that relies on the Jacobi identity of the Hamiltonian formulation, which we then use to obtain the tearing-layer parameter $\Delta '$  and the growth rate for the collisionless tearing mode. In Sec.~\ref{saturation} we use the conservation of a Casimir invariant  to obtain the nonlinearly saturated current  profile and compare it to that obtained by Rutherford.\cite{ruth73}  In Sec.~\ref{conclusions} we summarize and conclude.



\section{Model equations}
\label{model}

The model of  \cite{Fit04}  is given by the following equations:
\begin{eqnarray}
\label{e1}
\frac{\partial (\psi-{d_e^2}\nabla^2\psi)}{\partial t} + [\varphi,
\psi-{d_e^2}\nabla^2\psi] - {d_{\beta}}[\psi,Z] =0,\\
  \label{e2}
\frac{\partial Z}{\partial t} + [\varphi, Z] - {c_{\beta}}[v,\psi]
-{d_{\beta}}[\nabla^2\psi,\psi] =0,\\
\label{e3}
\frac{\partial\nabla^2\varphi}{\partial t} + [\varphi, \nabla^2\varphi] +
[\nabla^2\psi,\psi] =0,\\
  \label{e4}
\frac{\partial v}{\partial t} + [\varphi, v] -{c_{\beta}}[Z,\psi] =0.
\end{eqnarray}
Equation (\ref{e1}) is a reduced Ohm's law where the presence of 
finite electron inertia, which makes it possible for MR to take 
place, is indicated by the terms proportional to the electron skin 
depth $d_e$. Equations  (\ref{e2}), (\ref{e3}) and (\ref{e4}) are 
obtained from the electron vorticity equation, the  vorticity 
equation, and the parallel  momentum equation, respectively.

Considering a Cartesian coordinate system $(x,y,z)$ and taking $z$ as 
an ignorable coordinate,  the fields $\psi$, $Z$, $\varphi$ and $v$ 
are related to the magnetic field $\mathbf{B}$ and to the  
velocity field $\mathbf{v}$ by the relations $\mathbf{B}=\nabla\psi 
\times \hat{\mathbf{z}}+(B^{(0)}+c_{\beta}Z)\hat{\mathbf{z}}$ and 
$\mathbf{v}=-\nabla\varphi \times 
\hat{\mathbf{z}}+v\hat{\mathbf{z}}$, respectively.  Here $B^{(0)}$ is 
a constant guide field,  whereas $c_{\beta}=\sqrt{\beta/(1+\beta)}$ 
and $d_{\beta}=d_i c_{\beta}$ with $d_i$ indicating the ion skin 
depth.  For small $\be$, $\db\approx\rho_s$, the sonic Larmor radius.  
The ions are assumed to be cold, but electron pressure 
perturbations are taken into account and are given by 
$p=P^{(0)}+B^{(0)}p_1$, with $P^{(0)}$ a constant background 
pressure and  $p_1$ coupled to the magnetic field via the relation $p_1 
\simeq -c_{\beta} Z$.   Notice, here  the parameter 
$\beta$ is defined as $\beta=(5/3)P^{(0)}/{B^{(0)}}^2$, and  above  all the quantities are expressed in a dimensionless form 
according to the following normalization: $\nabla=a\nabla$, $t=v_A 
t/a$, $\mathbf{B}=\mathbf{B}/B_p$, where $a$ is a typical scale 
length of the problem, $B_p$ is a reference value for the poloidal 
magnetic field, and $v_A$ is the Alfv\'en speed based on $B_p$ and on 
the constant  density. Finally,   $[f,g]:=\nabla f 
\times \nabla g \cdot \hat{\mathbf{z}}$, for generic fields $f$ and 
$g$.



\section{Hamiltonian formulation}
\label{ham_form}

A desirable property for fluid models of the plasma is that the 
non-dissipative part of their equation of motion should admit a 
noncanonical Hamiltonian formulation \cite{mor82,mor98,mor05}. In short this 
means that it is possible to reformulate the ideal part of an 
$n$-field model as
\begin{equation}
  \frac{\partial \xi_i}{\partial t}=\{\xi_i,H\}, \qquad i=1, \cdots 
,n,\label{eq_HamDyn}
\end{equation}
where $\xi_i$ are suitable field variables, $H$ is the Hamiltonian 
functional and $\{ , \}$ is the Poisson bracket consisting of an 
antisymmetric bilinear form satisfying the Jacobi identity. 

The first task in the derivation of a noncanonical Hamiltonian 
formulation is to identify a conserved functional, usually the 
energy, that can serve as the Hamiltonian of the model.  If one 
considers, for instance, a square domain $\mathcal{D}$ in the $x-y$ 
plane with doubly  periodic boundary conditions, the four-field model 
(\ref{e1})--(\ref{e4}) admits the following constant of the motion:
  \begin{equation}
    \label{e:ham}
H=\frac{1}{2}\int_{\mathcal{D}}d^2 x
\, ({d_e^2 J^2} +{\vert\nabla \psi \vert^2} +{ \vert\nabla \varphi 
\vert^2}+{v^2} +{Z^2})
\end{equation}
with  $J=-\nabla^2 \psi$ indicating the parallel current density. The 
quantity $H$ represents the total energy of the system. The first 
term refers to the kinetic energy due to the relative motion of the 
electrons with respect to ions along the $z$ direction. The third and 
fourth terms account for the kinetic  energy, whereas the second 
and last terms account for the  magnetic energy.

Adopting $\psi_e=\psi-d_e^2 \nabla^2 \psi$, $U=\nabla^2\varphi$, $Z$, 
and $v$ as field variables, i.e.\
$\xi=(\psi_e,U,Z,v)$,  and (\ref{e:ham}) as Hamiltonian, it is 
possible to show that the model can indeed be cast in a noncanonical 
Hamiltonian form with the following Lie-Poisson bracket:
\bq
  \{F,G\}=\int d^2 x
\left(
U [F_{\xi},G_{\xi}]_U + \psi_e [F_{\xi},G_{\xi}]_{\psi_e} +
Z [F_{\xi},G_{\xi}]_{Z}
 + v [F_{\xi},G_{\xi}]_{v}\right)\,,
\eq
where
\begin{eqnarray}
[F_{\xi},G_{\xi}]_U &=& [F_U,G_U]\nonumber\\ 
{\ } \!\![F_{\xi},G_{\xi}]_{Z}&=&
[F_Z,G_U]+[F_U,G_Z]-d_{\beta}{d_e}^2[F_{\psi_e},G_{\psi_e}] 
\\
 &{\ }& \quad +c_{\beta}{d_e}^2([F_v,G_{\psi_e}]+[F_{\psi_e},G_v])
  -\alpha[F_Z,G_Z]-c_{\beta}\gamma [F_v,G_v] \nonumber\\
{\  } [F_{\xi},G_{\xi}]_{\psi_e}\!&=& [F_{\psi_e},G_U] +[F_U,G_{\psi_e}]
 -d_{\beta}([F_Z,G_{\psi_e}]+[F_{\psi_e},G_Z])  +c_{\beta}([F_v,G_Z]+[F_Z,G_v])\nonumber\\
{\ } \!\![F_{\xi},G_{\xi}]_{v}&=& [F_v,G_U]+[F_U,G_v]
+c_{\beta}{d_e}^2([F_Z,G_{\psi_e}]+[F_{\psi_e},G_Z] 
 -c_{\beta}\gamma([F_v,G_Z]+[F_Z,G_v]\nonumber\,,
\end{eqnarray}
with $\alpha=d_{\beta}+c_{\beta} {d_e}^2/{d_i}$, $\gamma= 
{d_e}^2/{d_i}$, and subscripts indicate functional differentiation. 
It is straightforward to show that Eq.~(\ref{eq_HamDyn}) with the 
Hamiltonian given in Eq.~(\ref{e:ham}) and the above bracket 
reproduces  equations (\ref{e1})-(\ref{e4}) of \cite{Fit04}. The Jacobi 
identity for the above Poisson bracket is readily established by 
applying the method described in Refs.~\cite{mor82,Thi00}.

Because the number of parameters is escalating, we record  their definitions here for later referral:
\bq
\db=\cb d_i\,,\quad
d=\sqrt{d^2_e + d^2_i}\,,
\quad
\ga= 
{d_e}^2/{d_i}\,,
\quad
\alpha=d_{\beta}+c_{\beta} {d_e}^2/{d_i}= \cb d^2/d_i\,.
\eq
The basic parameters of the model are $\cb,d_e$, and $d_i$, while $\db,\al,\ga$, and $d$ are useful shorthands. 



\section{Casimir invariants and bracket normal form}
\label{cassies}

Lie-Poisson brackets for noncanonical Hamiltonian systems are 
accompanied by the presence of Casimir invariants. A Casimir 
invariant is a functional that annihilates the Lie-Poisson bracket 
when paired with any other functional. That is,  a Casimir $C$ must 
satisfy
\begin{equation} \label{cas}
\{F,C\}=0,
\end{equation}
for every functional $F$. Thus, Casimir invariants constrain the 
nonlinear dynamics  generated by the Poisson bracket for any choice 
of Hamiltonian.

In order to identify the Casimirs of the four-field model we proceed 
in the following way. First, multiplying Eq. (\ref{e4}) times $d_i$ 
and adding it to Eq.~(\ref{e1}), we find
\begin{equation}  \label{lagD}
\frac{\partial D}{\partial t}+[\varphi,D]=0,
\end{equation}
where $D=\psi_e+d_i v$ is the ion canonical momentum.  
Equation (\ref{lagD}) indicates that the 
field $D$ is a Lagrangian invariant that is advected by the flow 
generated by the stream-function $\varphi$. The presence of this 
Lagrangian invariant also suggests that using $D$ as one of the 
variables will  simplify the Lie-Poisson bracket. Indeed, upon 
replacing $\psi_e$ with $D$ as field variable,  Eq.~(\ref{cas}) for 
the four-field model becomes
\begin{eqnarray}
\lefteqn{\{F,C\}=\int d^2 x \left( F_U [C_U,U]+F_D[C_U,D]+F_U[C_D,D] 
\right.}\hspace{12mm}\nonumber\\
& & \left. +c_{\beta} F_v[C_Z,D]  +c_{\beta} 
F_Z[C_v,D]+F_Z[C_U,Z]+F_U[C_Z,Z] \right. \nonumber\\
& & \left. -\alpha F_Z[C_Z,Z]-c_{\beta}\gamma 
F_v[C_v,Z]+F_v[C_U,v]+F_U[C_v,v] \right. \nonumber\\ 
& & \left. -\alpha F_v[C_Z,v]-\alpha F_Z[C_v,v] \right)=0,
\label{eq_simplrPB}
\end{eqnarray}
a simpler bracket.

For Eq.~(\ref{cas}) to be satisfied for any $F$, it is 
necessary for the coefficients of each of the functional derivatives 
of $F$ in (\ref{eq_simplrPB}) to vanish separately. This leads to the 
following system of equations for $C$:
\begin{eqnarray}
\label{e:C1}
[C_U,D]&=&0\,,\\
   \label{e:C2}
[C_U,U]+[C_D,D]+[C_Z,Z]+[C_v,v]&=&0\,,\\
  \label{e:C3}
-c_{\beta}[C_v,D]-[C_U,Z]+\alpha ([C_Z,Z]+[C_v,v])&=&0\,,\\
\label{e:C4}
c_{\beta}[C_Z,D]-c_{\beta}\gamma [C_v,Z]+[C_U,v]-\alpha [C_Z,v]&=&0.
\end{eqnarray}
The problem of finding the complete set of Casimir invariants is thus 
equivalent to solving the set of four coupled  equations 
(\ref{e:C1})--(\ref{e:C4}).  

Beginning with (\ref{e:C1}), a  functional integration shows that $C$ 
is of the form
\begin{equation}  \label{e:Cgen1}
C(U,D,Z,v)=\int d^2 x \,\big(
U \mathcal{F}(D)+g(D,Z,v)
\big)\,,
\end{equation}
where $\mathcal{F}$ and $g$ represent arbitrary functions of their 
arguments. The problem is now reduced to finding the  functions 
$\mathcal{F}$ and $g$. Considering next Eq.~(\ref{e:C2}), we see that 
this equation is automatically satisfied for any choice of $C$ with 
an integrand that depends only upon the field variables and not their 
spatial derivatives, and therefore imposes no constraints.
Substitution of  (\ref{e:Cgen1}) into (\ref{e:C3}) yields
\bq
\left(c_{\beta} g_{vv}+\alpha g_{ v D}\right) [v,D] 
+\left(c_{\beta}  g_{vZ} - \mathcal{F}'(D) 
+\alpha  g_{DZ}\right)[Z,D]=0\,,
\label{cas1}
\eq
where $'$ indicates derivative with respect to  argument and subscripts on $g$ denote partial derivatives. In (\ref{cas1}) the coefficients multiplying the   brackets 
`$[\ ,\ ]$' must vanish independently. These two relations  lead to 
  \begin{equation}   \label{e:relg1}
  c_{\beta} g_{v} +\alpha g_{D}=Z\mathcal{F}'(D)+K(D),
  \end{equation}
where $K$ an arbitrary function of $D$, and integration of (\ref{e:relg1}) by the method of characteristics gives
\bq
g(Z,v,D)=\frac{Z}{\al}\mathcal{F}(D)+\mathcal{K}(D)+\mathcal{G}(Z,v-\cb D/\al)\,,  
\label{cg}
\eq
where $\mathcal{K}'=K/\al$.  Similarly,  insertion of (\ref{e:Cgen1}) into  (\ref{e:C2}) yields  
\bq
\label{e:2g}
 \left(c_{\beta}  g_{Zv} -\mathcal{F}'(D) +\alpha  g_{ZD}\right) [D,v] 
+c_{\beta} \left( g_{ZZ} + \gamma  g_{vD}\right) [D,Z]  
+ \left(c_{\beta}\gamma  g_{vv} - \alpha   g_{ZZ}\right)[v,Z]=0\,,
\eq
which gives three equations, but only $g_{ZZ} + \gamma  g_{vD}=0$ provides new information.  Defining $V=v-\cb D/\al$, we see that $\mathcal{G}$ satisfies the wave equation
\bq
\mathcal{G}_{ZZ} - \frac{\cb\ga}{\al} \mathcal{G}_{VV}=0\,,
\eq
and thus has the general solution
\bq
\mathcal{G}=\sum_\pm
\mathcal{G}_\pm
\left(\pm\frac{1}{2 \alpha}\sqrt{\frac{c_{\beta}}{\gamma 
\alpha}}\left(D -\frac{\alpha}{c_{\beta}}v\right)-\frac{Z}{2 \alpha}\right)
\,.
\label{cpm}
\eq

Therefore, in light of (\ref{e:Cgen1}), (\ref{cg}), and (\ref{cpm}) we have the following four independent families of Casimir invariants:
\bqy
  C_1&=&\int d^2 x \left(U+\frac{Z}{\alpha}\right) \mathcal{F}(D)\,,
 \label{C1}
\\
C_2&=&\int d^2 x \, \mathcal{K}(D)\,,
\label{C2}\\
  C_\pm&=&\int d^2 x \,  
  \mathcal{G}_\pm
  \left(
  \pm\frac{d_i^2}{2\cb d^3d_e}  D 
  \mp\frac{d_i}{2c_{\beta}d d_e} v
  - \frac{d_i}{2\cb d^2} Z
\right)
 \label{Cpm}\,,
 \eqy
where we have scaled the arguments of $C_\pm$ for a reason that  will become apparent soon.

Knowledge of the functional dependence of the Casimirs  suggests a 
simplification of the Lie-Poisson bracket will occur if it is written 
in terms of  the new variables
  \begin{eqnarray}
  D&=&D\,,\label{e:nv1}\\
   \omp&=&U+\frac{Z}{\alpha}\,,\label{e:nv2}\\
  T_\pm&=& \pm \frac{d_i}{2\cb d^3d_e}\left(
  {d_i} D -  d^2 v \mp d d_e  Z
  \right)
\,, \label{e:nvpm}
   \end{eqnarray}
which possess  the inverse relations 
\begin{eqnarray}
  D&=&D\,, \label{e:nvi1}\\
U&=& \omp+ T_+ + T_-\,, \label{e:nvi2}\\
  Z&=&-\alpha(T_+ +T_-)\,, \label{e:nvi3}\\
v&=&\frac{d_i}{d^2} \, D  -\frac{\cb d_e d}{d_i}\left(T_+ -T_-\right) 
\,.
\label{e:nvi4}
  \end{eqnarray}
Indeed, in  the new variables,  the Lie-Poisson bracket reads
  \begin{eqnarray}
  \lefteqn{\{F,G\}=\int d^2 x \,
  (\omp[F_{\omp},G_{\omp}]+D([F_D,G_{\omp}]+[F_{\omp},G_D])}\hspace{32mm}
  \nonumber\\
  & &+{T}_-[F_{{T}_-},G_{{T}_-}]+{T}_+[F_{{T}_+},G_{{T}_+}])\,, 
  \label{clean_bkt}
  \end{eqnarray}
which is a bracket normal form that relies on the scaling used above.  The bracket in terms of the new variables   reveals its algebraic structure:  it  is identified as a sum  of direct product and semi-direct product parts \cite{mar_mor84,mor98,Thi00}  and, consequently,  the Jacobi identity follows from general theory.  Making use  of the variables suggested by the form of the Casimirs, the model  equations can be rewritten in the compact form
\begin{eqnarray}  
\label{e:lagrsys}
\frac{\partial D}{\partial t}&=&-[\varphi,D], \label{eq_Ddot}\\
\frac{\partial \omp}{\partial 
t}&=&-[\varphi,\omp]+ {d^{-2}}[D,\psi],
\label{eq_omdot}\\
\frac{\partial T_\pm}{\partial 
t}&=&-\left[\varphi_\pm,T_\pm\right],
\label{eq_Tpm}
\end{eqnarray} 
where for convenience we have defined
\bq
\varphi_{\pm}:=\varphi\pm \frac{\cb d}{d_e}\psi \,.
\eq
Note, the variable  $\omp$  plays the role of a  ``generalized'' vorticity, and our 
development reveals   the existence of the three Lagrangian 
invariants $D$, $T_+$ and $T_-$ associated with  the families of 
Casimirs $C_2$, $C_+$ and $C_-$, respectively. The existence of such 
invariants implies that the values of $D$, $T_+$ and $T_-$ are  
constant on the contour lines of $\varphi$, $\varphi_+$,   and 
$\varphi_-$, respectively.  By choosing ``top-hat''  functions for the free functions in the Casimirs $C_2$, $C_+$ and $C_-$, it follows that the area enclosed by the 
contour lines of the Lagrangian invariants remains constant. Notice 
also that $T_+$ and $T_-$ in the limit $\beta \rightarrow 0$ and $d_i 
\rightarrow \infty$  become  proportional to the Lagrangian invariants 
$G_{\pm}=\psi-d_e^2\nabla^2\psi\pm d_e \rho_s U$  of the two-field 
model derived in \cite{Sch94}. The family $C_1$ is of a different 
nature and one of the constraints imposed by it is that the total 
value of $\omp$ within an area enclosed by a contour line of $D$ 
remains constant.



\section{Equilibria}
\label{equilibria}

Equations governing the equilibrium of the FP model are most easily obtained by setting the time derivatives equal to zero in Eqs.~(\ref{eq_Ddot})-(\ref{eq_Tpm}) and solving for the the fields $D$, $\zeta$, and $T_\pm$. Alternatively, it is possible to derive equilibrium equations from a variational principle,  the existence of which is assured by the Hamiltonian nature of the equations.\cite{mor98} The construction of the variational principle is more laborious than the direct approach but the extra work is richly rewarded by the well-known benefits of variational principles. In particular, the variational principle provides a basis for studying the stability as well as the equilibria of the system.

For a system with a collection  of Casimirs, extrema of the free 
energy functional $ F=H+\sum  C$ are equilibria  of 
the system.  Such extrema  can be derived by setting the first 
variation $\delta F$ equal to zero.  If   $\xi_i, i=1,2,\dots$,  denotes 
 the   field variables of the system, this amounts to 
solving the   equations $H_{\xi_i}+\sum {C}_{\xi_i}=0$,  
where the subscript indicates functional derivative with respect to 
  $\xi_i$.  One  advantage of this variational approach is that for 
equilibria obtained as  extrema of $F$, the second variation of $F$ provides  
criteria  sufficient  for  stability. 


\subsection{General equilibria}
\label{GenEquil}

For the model here, the  free energy functional reads
\bqy
\label{e:freef}
F[\xi]&=&\int_{\mathcal{D}}d^2 x\left[
 \frac{d_e^2 \left(\nabla^2 \mathcal{L}\psi_e\right)^2}{2}
 + \frac{{\vert\nabla \mathcal{L}\psi_e \vert^2}}{2} 
+ \frac{\vert\nabla  \nabla^{-2} U \vert^{2}}{2}
+ \frac{v^2}{2} 
+ \frac{Z^2}{2}
+ \mathcal{K}(\psi_e +d_i v) 
\right.
\nonumber\\
&+&\left.
 \left(U+\frac{Z}{\alpha}\right)\mathcal{F}(\psi_e +d_i v) 
+\sum_\pm\mathcal{G}_\pm
  \left(
  \pm\frac{d_i^2}{2\cb d^3d_e}  D 
  \mp\frac{d_i}{2c_{\beta}d d_e} v
  - \frac{d_i}{2\cb d^2} Z
\right)
 \right]\,,
\eqy
where the linear operator $\mathcal{L}$ is defined by $\mathcal{L}^{-1}\psi=\psi-d_e^2 \nabla^2 \psi=\psi_e$.    Notice that the arguments of the functions present in  the Casimirs are of course much less compact when written in terms of  the variables $\xi=(\psi_e,U, Z,v)$. On the other hand, in terms of 
the variables $\up:=(D, \omp, T_+,T_-)$ the free energy functional  reads
\bqy
\label{e:freef2}
F[\up]
  &=&
\int_{\mathcal{D}}d^2 x
\left[
 \frac{c_{\beta}^2 d^4}{d_i^2}(T_+^2+T_-^2) 
 + \frac{D^2}{2d^2}-\frac{1}{2}\left(\omp+T_+ +T_-\right)
\nabla^{-2}
\left(\omp+T_+ +T_-\right) 
\right.
\nonumber\\
&-&
\left.
  \frac{1}{2} 
  \left(\frac{d_e}{d^2} {D} 
+ {c_{\beta}d} (T_+  -T_-)\right)
\mathcal{L}
  \left(\frac{d_e}{d^2} {D} 
+  {c_{\beta}d}  (T_+  -T_-)\right)
 \right.\nonumber\\
  &+&
\left.
\mathcal{K}(D)+\omp\mathcal{F}(D)+{\cal G}_+(T_+)+{\cal 
G}_-(T_-)\right]\,.
\eqy
Equation (\ref{e:freef2}) shows that in terms of the variables that  
are ``natural'' for the Casimirs, the expression for the Hamiltonian 
becomes  complicated. Unfortunately,   there exists no 
preferred set of variables in terms of which both the Hamiltonian and 
the Casimirs take a simple form. In order to obtain equilibrium 
solutions by means of the variational principle, it is 
convenient to choose the $\up$  variables that are  natural for the 
Casimirs, calculate the required functional derivatives of the Casimirs 
with respect to these variables,  and then use  the functional 
chain rule to obtain the functional derivatives of $H$ in terms of 
the variables $\xi$.  More specifically, by setting $\delta F=0$, the 
resulting equilibrium equations are given by
\bqy
\label{e:equ1}
F_{\omp}&=&H_{\omp}+\sum_{j} {C_j}_{\omp}=H_{\omp}+\mathcal{F}(D)=0,
\\
\label{e:equ2}
F_D&=&H_D+\sum_{j} {C_j}_D=H_D+\omp\mathcal{F}'(D)+\mathcal{K}'(D)=0,
\\
\label{e:equpm}
F_{T_\pm}&=&H_{T_\pm}+\sum_{j} {C_j}_{T_\pm}=H_{T_\pm}+{\cal G}_\pm '(T_\pm)=0.
\eqy
where the index $j$ ranges over the set $\{1,2,+,-\}$.
The functional chain rule then can be used to evaluate the functional derivatives of the Hamiltonian,
\begin{eqnarray*}
H_D&=&\frac{d_e^2}{d^2}\,H_{\psi_e}+\frac{d_i}{d^2}\,H_v,\\
H_{\omp}&=&H_U,\\
H_{T_\pm}&=&\pm c_{\beta} d_e d \, H_{\psi_e} + H_U - \alpha \, 
H_Z \mp \frac{\al d_e}{d}\, H_v.
\end{eqnarray*}
The functional derivatives $H_{\psi_e}$, $H_U$, $H_Z$ and $H_v$ can 
themselves easily be obtained from the Hamiltonian written in the form 
(\ref{e:ham}). The equilibrium equations 
(\ref{e:equ1})--(\ref{e:equpm}) are then given by
\bqy
\label{e:equ1b1}
-\varphi+\mathcal{F}(D)&=&0,
\\
\label{e:equ2b1}
-\frac{d_e^2}{d^2}\nabla^2 
\psi+\frac{d_i v}{d^2} +\omp\mathcal{F}'(D)+\mathcal{K}'(D)&=&0,\\
\label{e:equpmb1}
\mp c_{\beta} d_e d\, \nabla^2 \psi-\varphi-\alpha 
Z \mp \frac{\al d_e v}{d} +{\cal G}_\pm '(T_\pm)&=&0\,.
\eqy
These equations are expressed in a mixture of the $\xi$ and $\Upsilon$ variables. In order to simplify them we eliminate $\nabla^2\psi$ from the last two equations by using 
\[d_e^2\nabla^2\psi=\psi-D+d_i v.\]
Using this in Eq.~(\ref{e:equ2b1})-(\ref{e:equpmb1}), we find
\bqy
\label{e:equ2b2}
-d^{-2}\psi+\omp\mathcal{F}'(D)+\hat{\mathcal{K}}'(D)&=&0,\\
\label{e:equpmb2}
-\varphi_\pm \pm \frac{c_{\beta}  d}{d_e d_i}(d_i D-d^2 v\mp d d_e Z)+{\cal G}_\pm '(T_\pm)&=&0\,.
\eqy
where $\hat{\mathcal{K}}(D)=\mathcal{K}(D)+D^2/2d^2$. Equation (\ref{e:equpmb2}) may be simplified further by expressing $Z$ and $v$ in terms of the $\Upsilon$ variables using (\ref{e:nvi3})-(\ref{e:nvi4}). This leads to the following complete system of  equilibrium equations:
\bqy
\label{e:equ1b}
-\varphi+\mathcal{F}(D)&=&0,\\
\label{e:equpmb}
-\varphi_\pm+\hat{{\cal G}}_\pm '(T_\pm)&=&0,\\
\label{e:equ2b}
-d^{-2} \psi+\omp\mathcal{F}'(D)+\hat{\mathcal{K}}'(D)&=&0,
\eqy
where $\hat{{\cal G}}_\pm(T_\pm)={\cal G}_\pm(T_\pm)+\alpha^2 T_\pm^2$. One easily verifies that Eqs.~(\ref{e:equ1b})-(\ref{e:equ2b}) describe equilibrium states by substituting them into Eqs.~(\ref{e:lagrsys})-(\ref{eq_Tpm}).

Continuing our present approach of eliminating the $\xi$ variables would now lead us to express the $\psi$, $\varphi$, and $\varphi_\pm$ in terms of  integral operators acting on the $\Upsilon$ variables. Clearly this is undesirable. Instead, we note that the above four equations express a dependency between the six quantities $\nabla^2\psi$, $\nabla^2\varphi$, $\varphi$, $\psi$, $Z$ and $v$. It is thus possible in principle to use these equations to express four these quantities in terms of the remaining two. If we choose $\varphi$ and $\psi$ as the independent fields we will obtain a closed system of  equilibrium equations of the form
\begin{eqnarray}
\nabla^2 \psi&=&S(\psi,\varphi),
\label{eq_GSpsi}\\
\nabla^2\varphi &=&P(\psi,\varphi),
\label{eq_GSphi}
\end{eqnarray}
Equation~(\ref{eq_GSpsi}) is a generalized version of the Grad-Shafranov equation, whereas 
(\ref{eq_GSphi}) is an analogous equation that determines the equilibrium polarization. 

In order to calculate the form of the functions $S$ and $P$ we invert Eq.~(\ref{e:equ1b})-(\ref{e:equpmb}):
\bqy
\label{e:equ1c}
D&=&a(\varphi),\\
\label{e:equpmc}
T_\pm  &=& t_\pm(\varphi_\pm),\\
\label{e:equ2c}
\omp &=& d^{-2} \psi \, a'(\varphi)+b(\varphi),
\eqy
where $a$ and $t_\pm$ are the inverses of $\mathcal{F}$ and $\hat{{\cal G}}_\pm$ respectively and $b(\varphi)=-\hat{\mathcal{K}}'(a(\varphi))a'(\varphi)$. Solving these equations for $\nabla^2\varphi$ and $\nabla^2\psi$ then yields (\ref{eq_GSpsi})-(\ref{eq_GSphi}) with
\begin{eqnarray}
S(\psi,\varphi)&:=&\frac{\psi}{d_e^2 } -\frac{a(\varphi)}{d^2} 
- \frac{c_{\beta} d}{d_e}  \Big[ 
t_+(\varphi_+)-t_-(\varphi_-)\Big],
\label{eq_Sdef}\\
P(\psi,\varphi)&:=&b(\varphi)+t_+(\varphi_+)+t_-(\varphi_-)
 +a'(\varphi) \frac{\psi }{d^2}.
\label{eq_Pdef}
\end{eqnarray}
We complete the system by expressing the  two remaining unknowns $Z$ and $v$ in terms of $\varphi$ and $\psi$. From Eqs.~(\ref{e:nvi3}) and (\ref{e:nvi4}), using  (\ref{e:equ1c}), (\ref{e:equpmc}) there follows
\bqy
\label{eq_vfinl}
v(\psi,\varphi)&=&d_i\,a(\varphi)/d^2-c_\beta d_e d(t_+(\varphi_+)-t_-(\varphi_-))/d_i;\\
\label{eq_Zfinl}
Z(\psi,\varphi)&=&-\alpha(t_+(\varphi_+)+t_-(\varphi_-)).
\eqy 
 
We have thus shown  that solving the equilibrium equations amounts to 
solving the coupled system  of (\ref{eq_GSpsi}) and (\ref{eq_GSphi}) 
for the unknowns $\psi$ and $\varphi$. This requires making choices 
for  the free functions $t_{\pm}$, $a$,  and  $b$. If one is only interested in solving the equilibrium problem, one may determine these free functions directly from physical considerations and the relationship between these functions and those appearing in the variational principle may be ignored. These relationships become important, however, if one wishes to use the variational principle either to solve the equilibrium or the stability problem (the variational principle has well-known advantages both for numerical and analytic applications). In this case the functions $\mathcal{G}_{\pm},\mathcal{F}$, and $\mathcal{K}$, appearing in  the variational form may be determined in terms of $a$, $b$, and $t_\pm$ as described above. Variational treatments of two-fluid equilibria have been given by 
\cite{YoshiMaha,YoshiMahaOhsa,Goedb04}, and applications of variational principles to stability are discussed by \cite{YoshiMaha,YoshiMahaOhsa,HiroYoshHamri,HamriTora}.

We conclude this section by noting two difficulties with the equilibrium equations (\ref{eq_GSpsi})-(\ref{eq_GSphi}). The first is that these equations may become hyperbolic in the presence of strong flows. Recent analyses of this problem can be found in \cite{Goedb04,ItoRamoNaka}.  The second difficulty is that the right-hand sides are singular in the limits $d_e\ll L$ and $d_i \ll L$, where $L$ is a macroscopic scale length.\cite{YoshiMaha,YoshiMahaOhsa,SteinIshi06a} That is, for macroscopic equilibria the derivatives in Eqs.~(\ref{eq_GSpsi})-(\ref{eq_GSphi}) are multiplied by a small parameter, so that these equations form a stiff system. This has led to considerable grief, in particular for Field-Reversed Configuration (FRC) devices where a similar set of equations is encountered. Steinhauer has proposed a method for dealing with this problem that he named the ``nearby fluid'' approximation.\cite{SteinIshi06a,SteinGuo06b} In the following section we outline a similar approach to solving Eqs.~(\ref{eq_GSpsi})-(\ref{eq_GSphi}).

\subsection{Perturbative solution for macroscopic equilibria}

In order to deal with the singular nature of the equilibrium Eqs.~(\ref{eq_GSpsi})-(\ref{eq_GSphi}), we expand the fields in powers of the small parameters $d_e$ and $d_i$ and solve term by term. A byproduct of this procedure is the clarification of  the physical meaning of the profile functions.

We begin by considering the limit $d_e\rightarrow 0$.  For convenience we define
\[
\hat{t}_{\pm}\left(\pm \psi + \frac{d_e \varphi }{\cb d}\right)
:=d_e  \cb d\,  t_{\pm}\left(\varphi_{\pm}\right)
\]
and  expand this and other  profile functions in powers of $d_e$ according to
  \[
  \hat{t}_\pm(\phi)=\hat{t}_\pm^{(0)}(\phi)+d_e \hat{t}_\pm^{(1)}(\phi)+d_e^2 
\hat{t}_\pm^{(2)}(\phi)+\ldots \,.
\]
We then consider the equilibrium equations order by order in $d_e$.   From (\ref{eq_GSphi}) and (\ref{eq_Pdef}) we obtain at lowest order  $\hat{t}_+^{(0)}(\psi)=-\hat{t}_-^{(0)}(-\psi)$; using this result in 
(\ref{eq_GSpsi}) and (\ref{eq_Sdef}) we obtain to lowest order
$\hat{t}_+^{(0)}(\phi)=\hat{t}_-^{(0)}(\phi)=  {\phi}/{2}$.  To next order, (\ref{eq_GSpsi}) and (\ref{eq_Sdef}) yield $\hat{t}_-^{(1)}(-\psi)=\hat{t}_+^{(1)}(\psi)$; substituting this result in (\ref{eq_GSphi}) and (\ref{eq_Pdef}), we find the following equation for the vorticity
\[
\nabla^2\varphi = \hat{b}(\varphi)+ {a'(\varphi)}\psi/{d_i}^{2}
 -  {h(\psi)}/{\db}\,,
 \]
where $\hat{b}:=b+\varphi/\db^2$ is a profile function for the  vorticity and 
$h(\psi):=-2\hat{t}_+^{(1)}(\psi)$.

From the terms of order unity in (\ref{eq_GSpsi}) and (\ref{eq_Sdef}) we next find   
\begin{equation}
\nabla^2\psi=- a(\varphi)/d_i^2+h'(\psi)\varphi/\db+I(\psi),
\label{eq_GSdenaught}
\end{equation}
where $I(\psi)=\hat{t}_-^{(2)}(- \psi) - \hat{t}_+^{(2)}(\psi)$. Equation (\ref{eq_GSdenaught}) is the Grad-Shafranov equation, where the term proportional to $\varphi$ is the polarization current and  $I(\psi)$ describes the inductive current.

The parallel velocity may be obtained from  (\ref{e:equ1c}): to order $d_e^0$, $D=\psi-d_e^2\nabla^2\psi + d_i v= a(\varphi)$,  yields
\begin{equation}
d_i v+\psi=a(\varphi)\,.
\label{eq_ioncanmom0}
\end{equation}
Expanding (\ref{eq_Zfinl}) to order $d_e^0$,  gives 
\begin{equation}
Z+\varphi/d_\beta=h(\psi)\,.
\label{eq_frozin}
\end{equation}
 The 
sum $Z+\varphi/d_\beta$ represents  the electron stream-function. The 
fact that the electron stream-function is constant on surfaces of 
constant flux, as expressed by Eq.~(\ref{eq_frozin}), is a statement 
of the frozen-in property.  

We may carry out the limit $d_i\rightarrow 0$ in a similar way. From the ion momentum conservation Eq.~(\ref{eq_ioncanmom0}), we obtain $\psi=a(\varphi)$ showing that the electrostatic potential must be a flux function to lowest order. It is convenient to introduce $\Phi(\psi):=a^{-1}(\psi)=\varphi$ to denote the inverse of $a$. We also define the Alfv\'enic Mach number $M(\psi):=d\Phi/d\psi$. Note that  Eq.~(\ref{eq_frozin}) specifies that to lowest order, $h(\psi)=\Phi(\psi)/d_\beta$. In terms of these quantities, the vorticity equation shows that to lowest order,
\[\hat{b}(\varphi)=a(\varphi)a'(\varphi)/d_i^2-\varphi/c_\beta^2.\]
In order to  eliminate $\varphi$ from the Grad-Shafranov equation it is necessary to calculate the correction to the electrostatic potential. This is given by the vorticity equation,
\[M'(\nabla\psi)^2+M\nabla^2\psi=(M^{-2}-c_\beta^{-2})\varphi^{(2)}-h^{(1)}(\psi).\]
Note that the potential exhibits a singularity for $M=c_\beta$ corresponding to the sound-wave resonance. Lastly, after eliminating the electrostatic potential from Eq.~(\ref{eq_GSdenaught}) we recover the MHD version of the Grad Shafranov equation, 
\[(1-M^2)\nabla^2\psi-MM'(\nabla\psi)^2=\hat{I}(\psi).\]
In the following sections we present some explicit solutions of the equilibrium equations for simple profile functions.


\subsection{Quadratic Casimirs--dipole equilibria}
\label{qdipole}

The case of quadratic Casimir invariants is easily tractable.  Choosing
 \begin{equation}
\label{quadchoice}
\mathcal{K}(D)=\frac{A_D}{2} D^2, \qquad \mathcal{F}(D)=A_{\omp}D, \qquad 
{\cal G}_{\pm}(T_{\pm})=\frac{A_{\pm}}{2} T_{\pm}^2\, ,
\end{equation}
and following the steps of Sec.~\ref{GenEquil} leads to 
\bq
a(\varphi)= \frac{\varphi}{A_{\ze}}\,,\qquad
b(\varphi)= -\frac{\varphi}{A_{\ze}^2}
 \left(A_D + \frac{1}{d^2}\right)\,,
 \qquad
 t_{\pm}(\varphi_{\pm})= \frac{\varphi_{\pm}}{A_{\pm} +2\al^2}\,.
 \label{qpfl}
 \eq
Upon inserting (\ref{qpfl}) into (\ref{eq_GSpsi}) and (\ref{eq_GSphi}), we obtain
\bq
 \label{e:lin1b}
\nabla^2 \psi= S_{1}\psi + S_{2}\varphi
\qquad {\rm and }
\qquad
\nabla^2 \varphi= P_{1}\psi   + P_{2} \varphi  \,,
\eq
where $S_{1,2}$ and $P_{1,2}$ are constants that depend on  
$A_{\pm},A_{\ze},A_D$, and the parameters of the system.  
Note,  $S_{1,2}$ and $P_{1,2}$ are arbitrary  except that $S_2=-P_1$. Consequently,  Eqs.~(\ref{e:lin1b}) have a variety of solutions that are closely related to the double-Beltrami flows investigated by Yoshida et al.\cite{YoshiMaha,YoshiMahaOhsa}      
Specifically,  Refs.~\cite{YoshiMaha,YoshiMahaOhsa}  neglect electron inertial effects that are kept here, but they allow for more general geometry.  In general, (\ref{e:lin1b}) can be diagonalized resulting in two decoupled equations of the form
\bq
\nabla^2 \chi_i=-\la_i \chi_i\,,\quad i=1,2\,,
\eq
where $\la_{1,2}= -(S_1+P_2\pm\sqrt{(S_1-P_2)^2- 4P_1^2})/2$ and the $\chi_i$'s are linear combinations of $\varphi$ and $\psi$.  If a solution of this system is found, then one obtains $v$ and $Z$  as particular linear  combinations of  $\varphi $ and $\psi$  as described in Sec.~\ref{GenEquil}. 

\begin{figure}[htbp]
  \begin{center}
    \includegraphics[width=.4\linewidth]{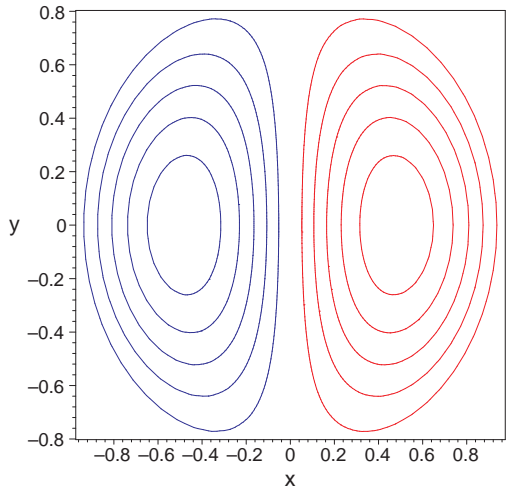}
  \end{center}
  \caption{$\chi$ contours for dipole solution of (\ref{dipole})}
  \label{fig:flux}
\end{figure}

Rather than describe the general solution, we give an example representative of the kinds of solutions that are possible.  We assume a circular domain of unit radius, adopt polar coordinates $(r,\theta)$, and adjusting the parameters $S_1, P_1$, and $P_2$ so that $\sqrt{\la_{1,2}}$ are zeros (possibly distinct) of the first order Bessel function, i.e.\  $J_1(\sqrt{\la_{1,2}})=0$.  With these assumptions, we obtain the solution
\bq
\chi_i(r,\theta)=A_i J_1(\sqrt{\la_i}r)\cos \theta\,,
\label{dipole}
\eq
where the $A_i$'s are constants and each of the $\chi$'s  has a dipolar structure like that  depicted in Fig.~\ref{fig:flux}.


\subsection{Homogeneous equilibria}
\label{homeq}

The quadratic Casimirs of (\ref{quadchoice}) also yield homogeneous equilibria, i.e.\ equilibria for which the linear dynamics has constant coefficients.   For this  choice,  the free energy functional  of (\ref{e:freef2}) can be written as follows:
\bq
F = \frac{1}{2}\int_{\mathcal{D}}d^2 x\left(\xi^T \hat{H}\xi + \up^T 
\hat{A} \up\right)\,,
\eq
where
\begin{equation}
\hat{H} =
\begin{pmatrix}
-\mathcal{L}\nabla^2& 0 & 0 & 0 \\
0 & - \nabla^{-2} & 0 & 0 \\
0 & 0 & 1 & 0 \\
0 & 0 & 0 & 1
\end{pmatrix}
\qquad {\rm and} \qquad
\hat{A}=
\begin{pmatrix}
A_D & A_{\ze}  & 0 & 0 \\
A_{\ze}  & 0 & 0 &   0 \\
0 & 0 & A_+ & 0 \\
0 & 0 & 0 & A_-   \end{pmatrix}\,.
\end{equation}
Recall $\xi=(\psi_e,U,Z,v)$ and $\up:=(D, \omp, T_+,T_-)$.  Equations 
(\ref{e:nv1})-(\ref{e:nvpm}) amount to $\up={\cal T}\xi$,  
where the   matrix ${\cal T}$ is given by
\begin{equation}
{\cal T}=\begin{pmatrix}
1 & 0 & 0 & d_i \\
0 & 1 & \frac{d_i}{ \cb d^2} & 0\\
  \frac{d_i^2}{2 \cb d_ed^3}  & 0&
  \frac{-d_i}{2 \cb d^2} &
\frac{-d_ed_i}{2 \cb d^3}  \\
 \frac{-d_i^2}{2 \cb d_ed^3}  & 0&
 \frac{-d_i}{2 \cb d^2}  &
\frac{d_ed_i}{2 \cb d^3} 
\end{pmatrix}.
\end{equation}
The free energy functional can then be written as a quadratic form:
\begin{equation}
F=\frac{1}{2}\int_{\mathcal{D}}d^2 x\left[\xi^T(\hat{H} +{\cal T}^T 
\hat{A} {\cal T})\xi\right]\,,
\end{equation}
whence the equilibrium equations  are  obtained upon  setting the 
functional derivatives $F_{\xi}$ to zero,
\begin{equation}
(\hat{H} +{\cal T}^T \hat{A} {\cal T})\xi=0.
\label{qequil}
\end{equation}
Here we have assumed the formal self-adjointness of the operators 
$\mathcal{L}\nabla^2$ and $\nabla^{-2}$.   Equation (\ref{qequil}) is a  linear 
homogeneous system of four equations with  the four unknowns 
$\psi_e$, $U$, $Z$, and $v$.  The equilibria treated in Sec.~\ref{qdipole} are solutions of this system, but of interest here are  the homogeneous equilibria  
\bq
\psi_0=  \als\, x\,, \quad
\varphi_0\equiv 0\,,  \quad
Z_0 =\alz\, x,\quad
v_0 =\alv\,x\,,
\label{e:equil1}
\eq
where the $\als$ is the Alfv\'{e}n speed, and $\alz$ and $\alv$   describe density and velocity shear, respectively.   These are clearly equilibrium solutions and can be related to the chosen Casimirs. Evidently, we are seeking solutions with $\nabla^2 \psi\equiv 0$ and $U\equiv 0$, so (\ref{qequil}) reduces to 
\begin{equation}
(\hat{{I}}^0_2+\mathcal{T}^T \hat{A}_3 \mathcal{T})\xi_3=0\,, 
\end{equation}
where $\xi_3:=(\psi,Z,v)$,  
\begin{equation}
\hat{I}^0_2 =
\begin{pmatrix}
0& 0 & 0 \\
0 & 1 & 0\\
0 & 0 & 1
\end{pmatrix}
\qquad {\rm and} \qquad
\hat{A}_3=
\begin{pmatrix}
A_D  & 0 & 0 \\ 
 0 & A_+ & 0 \\
 0 & 0 & A_-   \end{pmatrix}\,.
\end{equation}
This is a linear homogeneous system of three equations 
with unknowns $\psi$, $Z$,  and $v$.   The existence of non-trivial solutions requires  $\det (\hat{{I}}^0_2+\mathcal{T}^T \hat{A} \mathcal{T})=0$, 
which fixes a condition on $A_D$, $A_+$ and $A_-$ that assures $\psi, Z$, and $v$ are linearly dependent.  Thus,  they all can depend on $x$ and be proportional, consistent with (\ref{e:equil1}).  These equilibria  correspond to   uniform poloidal magnetic fields,  and 
to toroidal magnetic and velocity fields proportional to $x$.



\section{Normal forms for homogeneous equilibria}
\label{normal}

Here we work out the  linear canonical Hamiltonian form for the 
dynamics obtained by expansion about the equilibria of Sec.~\ref{homeq}. 
In terms of  the  $\up$ variables the equilibrium is 
\bq
D_0= \ald \,x, \quad
\omp_0=\alo x, \quad
T_{\pm 0}=\alpha_\pm \, x\,,
\eq
whence e.g.\ 
$\ald = \als +d_i\alv$ and (\ref{e:nvi4})  give $\als=\cb d d_e(\alp - \alm) + \ald(1- d_i^2/d^2)$. 
We obtain the Poisson bracket and Hamiltonian for the linear 
dynamics, obtain the dispersion relation, then we find a set of 
canonical variables, and discuss canonical transformations to normal 
forms.  En route we define the meaning of negative energy modes.


\subsection{Linear Hamiltonian form}

Denoting the linear variables with a tilde, i.e.\ $\omp=\omp_0 + 
\tilde \omp$, etc., and expanding the bracket  (\ref{clean_bkt}) 
gives the bracket for the linear dynamics (see \cite{mor98})
\begin{eqnarray*}
\{F,G\}_L=\int d^2x\left(
  \omp_0\left[\fw,\gw\right]+
D_0 \left(\left[\fd,\gw\right] + \left[\fw,\gd\right]\right)
+ T_{\pm0} \left[\fpm,\gpm\right] \right),
\end{eqnarray*}
where a sum over the $\pm$ terms is implied.
Upon integration by parts this bracket becomes
\begin{eqnarray}
\{F,G\}_L=- \int d^2x\left[
\alo\fw \frac{\partial}{\partial y} \gw+\ald \left(\fd\py \gw  + 
\fw\py \gd \right)
 + \alpha_\pm  \fpm\py \gpm  
\right] .
\label{linbkt}
\end{eqnarray}
The above bracket together with the following quadratic Hamiltonian:
\bq
H_L=\frac1{2}\int d^2x\left(\tilde\psi_e \tilde J
+|\nabla \tilde\phi |^2 + \tilde{v}^2 + \tilde{Z}^2 +
  A_D \tilde{D}^2 + A_\pm\tilde{T}_\pm^2
\right)
\label{hl}
\eq
where
\bq
A_D= 
\frac1{d^2}\left(\frac{\als}{\ald} -1 \right) 
\quad
{\rm and}
\quad
A_{\pm}=c_{\beta}\left(\pm\frac{d}{d_e}\,\frac{\als}{\al_{\pm}}\ 
- 2 \frac{c_{\beta}d^4}{d_i^2} \right),
\eq
when written entirely in terms of the variables $(\tilde{\omp}, 
\tilde{D}, \tilde{T}_+,\tilde{T}_-)$, yield the linearized equations 
of motion in Poisson bracket form.

Unlike the nonlinear semi-direct product bracket of (\ref{clean_bkt}), the linear bracket of (\ref{linbkt}) can be brought into direct product form by the transformation
\bq
\bar{D}= -\alo \tilde{D} +\ald \tilde{\ze}\,,
\label{linbktDPtra}
\eq
which yields the convenient form
\begin{eqnarray}
\{F,G\}_L&=&- \int d^2x\left[
\alo\fw \frac{\partial}{\partial y} \gw+\al_{\bar D}  \frac{\delta F}{\delta \bar D} \py 
\frac{\delta G}{\delta \bar D} +
 \alpha_\pm  \fpm\py \gpm 
\right] ,
\label{linbktDP}
\end{eqnarray}
where $\al_{\bar D}:=-\ald^2\alo$.


\subsection{Canonical coordinates}
\label{cancor}

Because the equilibrium equations do not depend explicitly on $x$ and 
$y$,  we Fourier expand $\tilde{\omp}$ as
\bq
\tilde{\omp}(x,y,t)=\sum_{k_x,k_y=-\infty}^{\infty} 
\omp_{k_x,k_y}(t)\,  e^{-i(k_xx+k_yy)}
\label{ft}
\eq
and similarly for  $\bar{D}, \tilde{T}_+$, and $\tilde{T}_-$.  For 
convenience we suppress the sum over $k_x$   and set $k_y=k$.  The 
variable $k_x$ will only appear in the combination $\kt^2:=k_x^2 
+k_y^2$.  It is not difficult to prove (see e.g.\ 
\cite{Gardner_kdv}) the  following general functional derivative 
relationship:
\bq
\fw= \st \left(\fw\right)_{k}\, e^{-iky}
=\frac1{2\pi} \st \fomk \, e^{-iky}\,,
\label{fti}
\eq
where $F[\ze]=\bar{F}[\ze_k]$.  Note, in the above expression there is an extra factor  of $2\pi$ that occurs in the  denominator  of the suppressed  sum on $k_x$, but this factor is compensated by a factor of $2\pi$ that accompanies a suppressed sum in the Hamiltonian.

Upon inserting (\ref{fti}) and similar relations for the other 
variables into the bracket of (\ref{linbkt}) gives
\bqy
\{\bar F,\bar G\}
&= & \su \frac{ik}{2\pi} \left[ 
\alo \left(
\fok \gomk - \fomk \gok 
\right)  
+  \al_{\bar D} \left(
 \frac{\delta \bar F}{\delta \bar D_k} \frac{\delta \bar G}{\delta \bar D_{-k}} 
 -  \frac{\delta \bar F}{\delta \bar D_{-k}} \frac{\delta \bar G}{\delta \bar D_k}
 \right)\right.
 \label{Fbkt}\\
&+& \left. \alp \left(\fptk \gptmk - \fptmk \gptk \right)
+   \alm  \left(\fmtk \gmtmk - \fmtmk \gmtk \right)
\right]\nonumber\,, 
\eqy
where  in order to facilitate the transformation to canonical 
variables the positive and negative values of $k$  are assumed to be 
independent and the sum is  written over only positive values of $k$.

A set of real valued canonical variables is given by
\bqy
q_k^{(1)}&=&-\sqrt{\frac{\pi}{k |\alpha_{\bar D}|}}
\left(\bar{D}_k + \bar{D}_{-k}\right) \,,
\quad\quad \quad\ \ \ \
p_k^{(1)}= i\sqrt{\frac{\pi}{k |\alpha_{\bar D}|}} \left(\bar{D}_k - 
\bar{D}_{-k}\right) \,,
\nonumber\\
q_k^{(2)}&=&\sqrt{\frac{\pi}{k \alpha_{\omp}}}
\left(\omp_k + \omp_{-k}\right) \,,
\quad\quad \quad\ \ \ \
\quad p_k^{(2)}= i\sqrt{\frac{\pi}{k \alpha_{\omp}}} \left(\omp_k - 
\omp_{-k}\right) \,,
\nonumber\\
q_k^{(3)}&=&\sqrt{\frac{\pi}{k \alpha_+}}
\left(\ptk + \ptmk\right)
  \,,
\quad\quad \quad p_k^{(3)}= i\sqrt{\frac{\pi}{k \alp}} 
\left(\ptk - \ptmk \right)
\,,
\nonumber\\
q_k^{(4)}&=&\sqrt{\frac{\pi}{k \alm}}
\left(\mtk+ \mtmk\right)
  \,,
\quad\quad \quad
p_k^{(4)}=  i \sqrt{\frac{\pi}{k \alm}}  \left(\mtk - \mtmk\right) \,,
\eqy
with the inverse transformation  
\bqy
\bar{D}_k&=&-\frac1{2}\sqrt{\frac{k |\alpha_{\bar D}|}{\pi}}
\left(q_k^{(1)} +i  p_k^{(1)}\right)\,,
\quad\quad\quad
\bar{D}_{-k}=-\frac{1}{2}\sqrt{\frac{k|\alpha_{\bar D}|}{\pi}}
\left(q_k^{(1)} - i p_k^{(1)}\right)\,,
\nonumber\\
\omp_k&=&\frac1{2}\sqrt{\frac{k \alo}{\pi}}
\left(q_k^{(2)} -i  p_k^{(2)}\right)\,,
\quad\quad\quad
\omp_{-k}=\frac1{2}\sqrt{\frac{k \alo}{\pi}}
\left(q_k^{(2)} + i p_k^{(2)}\right)\,,
\nonumber\\
\ptk&=&\frac1{2}\sqrt{\frac{k \alp}{\pi}}
\left(q_k^{(3)} - i  p_k^{(3)}\right)
\,,
\quad\quad\quad
\ptmk=\frac1{2}\sqrt{\frac{k \alp}{\pi}}
\left(q_k^{(3)} + i p_k^{(3)}\right)\,,
\nonumber\\
\mtk&=&\frac1{2}\sqrt{\frac{k \alm}{\pi}}
\left(q_k^{(4)} - i  p_k^{(4)}\right)
\,,
\quad\quad\quad
\mtmk=\frac1{2}\sqrt{\frac{k \alm}{\pi}}
\left(q_k^{(4)} + i  p_k^{(4)}\right)\,.
\eqy
Above we have assumed that $\alo,\alp$,  and $\alm$  are positive;  in light of its definition, $ \al_{\bar D}$ is negative and there is an intrinsic parity difference between the $\bar D_{\pm k}$  and  the $\ze_{\pm k}$ degrees of freedom.  This is a fundamental property of linearized semi-direct product brackets.  If $\al_{\bar D}$ is positive, then $\alo$ is negative, and one must alter the $\zeta_{\pm k}$ transformations.    In  general, if any of the $\al$'s are negative, then a suitable canonizing transformation is obtained by inserting an absolute value inside the square root and replacing the corresponding $q_k$ by minus $q_k$.

The Hamiltonian corresponding to (\ref{Fbkt}) is
\bqy
H_L&=&{2\pi}\su 
 \frac{ a_1}{\alo^2} |\bar{D}_k|^2 - \frac{a_1}{\alo}\sqrt{{|\alpha_{\bar D}|}/{\alo}}
 \left(
 \bar{D}_k\ze_{-k} + \bar{D}_{-k}\ze_{k}
 \right)  
+ \left(a_2+ a_1{|\alpha_{\bar D}|}/{\alo}\right)|\ze_k|^2   
  \nonumber\\
  &+&  a_3 |{{T}_+}_k|^2  +a_4|{{T}_-}_k|^2 + a_5\left(
 {{T}_+}_k {{T}_-}_{-k} +{{T}_+}_{-k}{{T}_-}_k\right) 
+  \frac{a_6}{\alo}
 \left(
  \bar{D}_k {{T}_+}_{-k} + \bar{D}_{-k}{{T}_+}_{k} \right)
 \nonumber\\
 &-&
 \frac{a_6}{\alo}   \left(
  \bar{D}_k {{T}_-}_{-k} + \bar{D}_{-k}{{T}_-}_{k} \right)
 +  \left(a_2 -\frac{a_6}{\alo} \sqrt{{|\alpha_{\bar D}|}/{\alo}}\right)
 \left( \ze_k {{T}_+}_{-k} +\ze_{-k} {{T}_+}_{k}\right)
 \nonumber\\
 &+& 
  \left(a_2 +\frac{a_6}{\alo} \sqrt{{|\alpha_{\bar D}|}/{\alo}}\right)
 \left( \ze_k {{T}_-}_{-k} +\ze_{-k} {{T}_-}_{k}\right)\,.
\label{hlFT}
\eqy
In term of the canonical variables this Hamiltonian becomes
\bqy
H_L&=& \frac1{2}\su  \sum_{i,j=1}^4 k \left(M_{ij} 
p_k^{(i)} p_k^{(j)}  +  K_{ij}q_k^{(i)} q_k^{(j)} \right)
  \,,
  \label{hamL}
\eqy
where the symmetric matrices  $M$ and $K$ are  given by
\bqy
&{\ }&M_{11}= K_{11} =  \frac{\aalbd}{\alo^2} a_1\,,\quad\
M_{22}=  K_{22}= \frac{\aalbd}{\alo^2} a_1 + \frac{a_2}{\alo}  \,,\quad\
M_{33}= K_{33}= \alp a_3\,,
\nonumber\\
&{\ }&M_{44} =  K_{44}=  \alm a_4\,,
\quad
M_{12}= - K_{12}=  - \frac{\aalbd}{ \alo^2} a_1\,,\quad\
M_{13}=-K_{13}=     \frac{\sqrt{\aalbd\alp}}{\alo} a_6\,,
\nonumber\\
&{\ }&M_{14}=- K_{14}= -\frac{\sqrt{\aalbd\alm}}{\alo} a_6\,,\quad\
M_{34}=K_{34}=   \sqrt{\alp\alm} a_5\,,
\\
&{\ }&M_{23}=K_{23}=  \sqrt{\alp\alo} a_2\ -   \frac{\sqrt{\aalbd\alp}}{\alo} a_6\,,
\quad\
M_{24}=K_{24}=   \sqrt{\alo\alm} a_2 + \frac{\sqrt{\aalbd\alm}}{\alo} a_6\ \,,
\nonumber
\eqy
with
\bqy
a_1&=&\frac1{d^2} +A_D - \frac{d_e^2}{d^4(1+d_e^2 \kt^2)} 
= \frac{\als}{\ald d^2}   -\frac{d_e^2}{d^4(1+d_e^2 \kt^2)}
\,,\quad\ \nonumber\\
a_2&=&\frac1{\kt^2}
\,,\quad\ \nonumber\\
a_{3,4}&=&\frac1{\kt^2}+ A_{\pm}
+ \frac{2 c_{\beta}^2d^4}{d_i^2} 
- \frac{c_{\beta}^2 d^2}{1+d_e^2 \kt^2} 
= \frac1{\kt^2} \pm \frac{\cb d}{d_e}\frac{\als}{\al_{\pm}}
- \frac{c_{\beta}^2 d^2}{1+d_e^2 \kt^2} 
\nonumber\\
a_5&=& \frac1{\kt^2} + \frac{c_{\beta}^2 d^2}{1+ d_e^2 \kt^2}
\nonumber\\
a_6&=&\frac{c_{\beta}d_e}{ (1+ d_e^2 \kt^2)d}\,.
\eqy
Note, the matrices $M$ and $K$ have commuting and anti-commuting parts, which can be traced to the intrinsic parity difference mentioned above.


\subsection{Stability, signature, and normal forms}
\label{stabsignf}

One could proceed directly by linearizing the system of 
Eqs.~(\ref{e1})--(\ref{e4}) about the equilibrium (\ref{e:equil1}) 
and obtain a system for the linear dynamics.  Because we are using 
variables indexed  by $k$ and $-k$ as independent variables,  we 
obtain the combined linear system
\bq
\dot{\tilde\xi}_k=  \mathcal{R}_k\cdot  \tilde\xi_k\quad\quad
{\rm and}
\quad\quad
\dot{\tilde\xi}_{-k}=  \mathcal{R}_{-k}\cdot  \tilde\xi_{-k}
\,,
\label{dot1}
\eq
where ${\tilde\xi}_{\pm k}$ is a four dimensional vector and $\mathcal{R}_{\pm k}$ is a $4\times 4$ matrix, for each value of $k=1,2,\dots$. 
This means we have an eight-dimensional system for each value of $k$, 
and upon assuming $\tilde\xi\sim\exp(i\om t)$ we obtain the 
dispersion relation from
\bqy
\det
  \begin{pmatrix}
  i\om{I}_4 - \mathcal{R}_k & \ \ O_4\ \
\\
\ \ O_4\ \  & i\om{I}_4 -\mathcal{R}_{-k}
  \end{pmatrix}
  &=&\det(i\om{I}_4 - \mathcal{R}_k)\cdot\det(i\om{I}_4 - {\cal 
R}_{-k})\nonumber\\
  &=&\det(-\om^2{I}_4 + \mathcal{R}_k^2)=
  0\,.
  \label{8desr}
\eqy
where $O_4$ is a $4\times 4$ matrix of zeros,  ${I}_4$ is the 
$4\times 4$ identity matrix, and use has been made of $\mathcal{R}_{-k}= 
-\mathcal{R}_{k}$ which is easily verified for the case at hand. 
Alternatively, one can assume  $(q_k^{(i)},p_k^{(i)})= 
(\hat{q}_k^{(i)}\exp(i\om t),\hat{p}_k^{(i)}\exp(i\om t))$, and 
obtain the dispersion relation from Hamilton's equations  in the form
\bq
i\om \hat{q}_k^{(i)}= \sum_{j} k M_{i j}\, \hat{p}_k^{(j)} \quad\quad
{\rm and} \quad \quad
i\om  \hat{p}_k^{(i)}= -\sum_{j} k K_{i j}\, \hat{q}_k^{(j)}\,,
\eq
whence one obtains the dispersion relation as $\det(\om^2{\cal I}_4- 
k^2 MK)= 0$, a relation equivalent to (\ref{8desr}) with $MK=\mathcal{R}_k^2$.  Because of this special form,  the dispersion  relation can be factored,  and reduced to a $4\times 
4$ determinant that provides the frequencies for both positive and negative $k$. 
Thus the symmetry $\mathcal{R}_{-k}= -\mathcal{R}_{k}$  allows for a 
simplification that is manifested  by the special form of the 
Hamiltonian of (\ref{hamL}).  We note that this special form occurs for all the basic fluid and plasma models, because they have real variable Hamiltonian form.

For convenience,  we introduce the dimensionless variables
\bqy
&{\ }&   v_A=\als   \,, \quad \kat= \kt d_e\,,
\quad r= \frac{d_e}{\db}\,, \quad N=\frac{\om r}{k {v}_A} 
     \nonumber\\
&{\ }&
   \delta = 
\cb r = \frac{d_e}{d_i}=\sqrt{\frac{m_e}{m_i}}
\,, \quad   s = \frac{\alv \db r}{{v}_A}\,,
     \quad   \nu =\frac{\alz\db r}{{v}_A},
\eqy
in terms of which we derive the following dispersion relation:
\bq
\label{dispn}
(1+\kat^2)N^4 -\nu N^3-\left(\delta^2  +\delta^2 \kat^2 +\delta  s + r^2+ 
  \kat^2 \right)N^2+ \nu r^2 N 
+r^2 \delta(s+\delta)=0 \,,
\eq
{}From the form of 
(\ref{dispn}) it is clear that $N$ is a function of $\kat^2$ alone; moreover,  it can be rewritten as 
\bq
\kat^2=-\frac{(N^2-N_r^2)(N-N_+)(N-N_-)}
{N^2(N^2-N_{\de}^2)}\,,
\label{kat}
\eq
where
\bq
N_{\pm}=\frac{\nu \pm \sqrt{\nu^2 +4\de(\de + s)}}{2}\,,\quad N_r={r}\,,
\quad {\rm and} \quad N_{\de}=\sqrt{1+\de^2}\,.
\eq
Analysis of the dispersion relation reveals that the four roots 
correspond to two that are Alfv\'{e}n-like, two that are  a combination of a  
drift-like wave  that arises from density shear $\nu$ a  Kelvin-Helmholtz-like wave  that arises from  the parallel velocity shear $s$. 

For example, if one sets $\de=0$ and $\nu=0$, then (\ref{kat}) yields in dimensional variables 
$\om =  \pm k v_A \sqrt{(1+\kt^2\db^2)/(1+\kt^2d_e^2)}$. 
The square  root in  this  dispersion relation displays the slippage of flux, which comes from two sources, a numerator that is dependent on the 
electron temperature through $\db$ and a denominator that is 
dependent on electron inertia through $d_e$.  Both of the terms in 
the square root break the MHD frozen-in condition and for $d_e=\db=0$ one 
recovers the Alfv\'{e}n wave dispersion relation.  For $d_e=0$ and 
$\db\neq 0$ this  dispersion relation reduces to that for a 
version of the `kinetic'  Alfv\'{e}n wave, while for large $k$ one 
obtains  the phase velocity $\om/k\sim  v_A\db/d_e= \sqrt{T_e/m_e}$, the electron thermal speed.  For $\db=0$ and large $k$,   the lower hybrid frequency 
$\om\sim  v_A/d_e= v_A  \om_{pe}/{c}= {eB}/c\sqrt{m_em_i}$ is obtained.  Similar limits reveal the 
presence of drift-waves associated with $\nu$ and  Kelvin-Helmholtz modes 
associated with $s$.  For example, if we set $\de=0$ and  suppose $N$ is small for   
large wave-numbers, then  (\ref{kat}) gives the drift-wave dispersion relation, 
$\om= {kv_*}/(1+\kt^2\db^2)$, where $v_*=\db \alz$.
 
Equation (\ref{kat}) is convenient for obtaining stability criteria, by examination of the  zeros and divergences of its right hand side, and by noting that it asymptotes to unity for large $|N|$.  Because $\de$ and $r$ are always positive, the sign of the divergence at $N=0$ is governed by $\de +s$, and this sign distinguishes two cases.  The  {\it first case}, $\de +s>0$, corresponds to positive or weakly negative parallel velocity shear.  It is convenient to define the central band  by $\mathcal{C}_{\de}:=\{N| |N|< N_{\de}\}$, which is bordered by the divergences, and the set of   frequencies at which zeros occur, $\mathcal{N}:=\{N_+,N_-,N_r,-N_r\}$.  If any two elements of $\mathcal{N}$ are contained in $\mathcal{C}_{\de}$, then the system is stable.  If three elements of $\mathcal{N}$ are contained in $\mathcal{C}_{\de}$,   then the system is stable, and if all four of the elements of $\mathcal{N}$ are contained in $\mathcal{C}_{\de}$, then the system is stable.   The {\it second case},  $\de +s<0$, corresponds to strong negative velocity shear.  If $\nu^2 + 4\de(\de + s)>0$, then the system is unstable for large enough $\kat$, but always possesses two stable modes.  In the case of very strong negative velocity shear $\nu^2 + \de(\de + s)<0$, the  set of zeros becomes $\mathcal{N}:=\{N_r,-N_r\}$, and the system is unstable for all $\kat$, but again always possesses two stable modes.

\begin{figure}[htbp]
  \begin{center}
    \includegraphics[width=.4\linewidth]{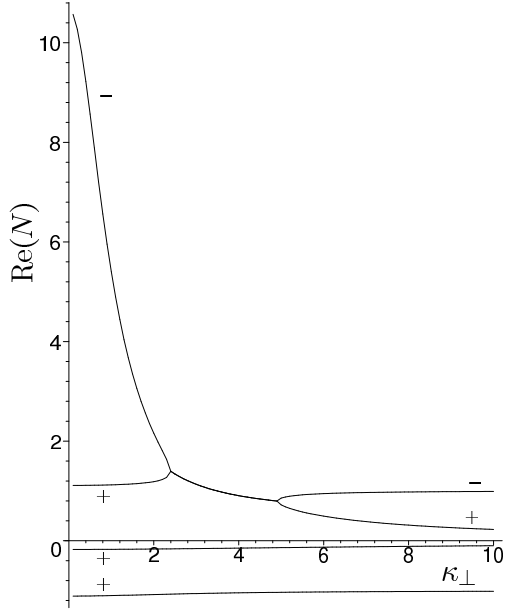}
  \end{center}
  \caption{Solution of the dispersion relation of (\ref{dispn}). The 
real part of $N$ is plotted vs.\ $\kat$  for $r=1.11$, 
$\nu=10.5$, $\de =0.10$, $s= 18.5$, which corresponds to positive  parallel velocity shear with $\de +s>0$.  The energy signatures of the modes are indicated.}
  \label{fig:d1}
\end{figure}

\begin{figure}[htbp]
  \begin{center}
    \includegraphics[width=.4\linewidth]{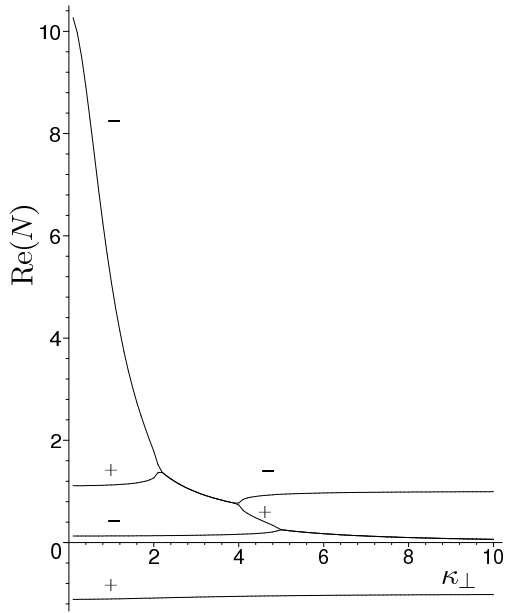}
  \end{center}
  \caption{Solution of the dispersion relation of (\ref{dispn}). The 
real part of $N$ is plotted vs.\ $\kat$ for   $r=1.11$, 
$\nu=10.5$, $\de =0.10$,  $s=-13.4$,   which corresponds to strong negative  parallel velocity shear with $\de +s<0$.  The energy signatures of the modes are indicated.}
  \label{fig:d2}
\end{figure}

In Figs.~\ref{fig:d1} and \ref{fig:d2} two   solutions of  (\ref{kat}), with the  real part of  $N$ versus  $\kat$,  are plotted.  
  In Fig.~\ref{fig:d1}  the parameters $r=1.11$, $\nu=10.5$, 
$\de =0.10$ , and $s= 18.5$ are used, while in Fig.~\ref{fig:d2}  the same values except $s=-13.4$ are  used. 
For small $\kat$, the uppermost curve 
corresponds to the drift-shear wave that has $N=N_+$ at $\kat=0$.  The second to upper is the Alfven wave, which has $N=N_r$ at $\kat=0$ corresponding to $\om/k=v_A$, and the lowermost curve is  its negative counterpart.  The remaining wave is a drift-shear mode with  $N=N_-$ at $\kat=0$. For larger $\kat$ there exist regions of instability.

For the stable degrees of freedom of Hamiltonian systems there exists 
a special form, a so-called normal form, to which all such systems 
can be mapped by a canonical transformation, 
$(q,p)\longleftrightarrow (Q,P)$.  The algorithm for  this, which 
uses the real and imaginary parts of the linear eigenvectors,   was first proven in total generality in  \cite{Will}. (See e.g.\   \cite{Meyer} for a more recent  source for a 
version of the algorithm and \cite{MK90} where a plasma  example is 
worked out.)\ \  For the stable modes  this 
normal form Hamiltonian is given by
\bq
H_L'=\frac1{2}{\sum_k}'\sum_{i=1}^4 \si^{(i)}_k\,  \om_k^{(i)} 
\left({P_k^{(i)}}^2 + {Q_k^{(i)}}^2\right)\,,
\label{StNoFo}
\eq
where the prime on the sum means the $k_y$ and $k_x$ values for the unstable  modes are absent (recall the suppressed sum on $k_x$) and the 
frequencies $ \om_k^{(i)}>0$.  This Hamiltonian is merely that for a 
collection  of simple harmonic oscillators, except for the presence 
of the signature  $\si_k^{(i)}\in\{1,-1\}$.   Modes with signature 
$\si_k^{(i)}=-1$ oscillate,  but are negative energy modes.

The signatures of the modes can be obtained by inserting the eigenvectors into the Hamiltonian (\ref{hlFT}). If $\up_k^{(i)}$ is the eigenvector associated with the mode indexed by $i$ and $k$, then its signature is given by the sign of $H_L={\up_k^{(i)}}^*\hat{H}_L \up_k^{(i)}$, where $\hat{H}_L$ is the matrix of the bilinear form (\ref{hlFT}). To determine the signature of all modes, it is only necessary to do this for  $\kat \rightarrow \infty$ and $\kat \rightarrow 0$.

From (\ref{kat}) it follows for $\kat\rightarrow \infty$ that  $\om\kat /k\sim \pm v_A\sqrt{\de(\de +s)/(1+\de^2)}$. For these two modes in this limit, $(\ref{hl})$ is dominated by contributions from the  terms $|J|^2$, $|\nabla\varphi|^2$, $v^2$, and $D^2$, giving
\bq
H_L\sim  \frac{\kat^2|\varphi_k|^2 }{d_e^2}\,,
\quad\quad\quad N\rightarrow 0\,,\quad \kat \rightarrow \infty\,, 
\label{sig1}
\eq
and thus they both have positive signature.  Note, because $s +\de <0$,  these modes are unstable for Fig.~\ref{fig:d2} and (\ref{sig1}) does not apply.  For the remaining two modes  the energy in this limit behaves as 
\bq
H_L \sim \frac{\kat^4|\varphi_k|^2(1+\de^2)^2}{r^2d_e^2(N_{\de}\mp N_+)(N_{\de}\mp N_-)}
\,,\quad\quad\quad N\rightarrow \pm N_{\de}\,,\quad   \kat \rightarrow \infty\,.
\label{sig}
\eq
From (\ref{sig}) if follows that  for the examples of Figs.~\ref{fig:d1} and \ref{fig:d2}, the   mode approaching $N_{\de}$ is negative while that approaching $-N_{\de}$ is positive.

In the limit $\kat\rightarrow 0$ the modes have the values $\pm N_r$ and $N_{\pm}$,  plus corrections to these values of order $\kat^2$. For the Alfven waves that emerge from $\pm N_r$ the energy vanishes to leading order, reflecting the vanishing of line-bending in this long wavelength limit, but proceeding to order $\kat^2$ gives  positive energies for all  the stable  Alfven waves  of Figs.~\ref{fig:d1} and \ref{fig:d2}.  Note, for Fig.~\ref{fig:d1},  the signature of the negative Alfven wave  $-N_r$ is consistent with the result of the $\kat\rightarrow \infty$ calculation: because this mode does not traverse zero or suffer a bifurcation to instability, it cannot change signature.   Similarly we obtain that  the drift-shear mode emerging from  $N_{+}$ has negative energy for the values of both Figs.~\ref{fig:d1} and \ref{fig:d2}, while the $N_{-}$ mode has positive energy for Fig.~\ref{fig:d1} and negative energy  for  Fig.~\ref{fig:d2}.  As $s$ goes from $18.5$ to $-13.4$, its frequency goes through zero and the mode changes signature.

In Fig.~\ref{fig:d1}  there is an unstable gap  that  lies  
between $2.39 \leq \kat\leq  4.91$.  At the value   $\kat\approx 2.39$, we have a `collision' 
where the frequencies of the Alfven wave and the upper drift-shear wave match and 
a transition to instability occurs.  This transition is one of two 
types that occur in Hamiltonian systems;  the other occurs at zero 
frequency as is the case for ideal MHD instabilities of static 
equilibria.  A necessary condition for the transition to instability 
at nonzero frequency is that one of the modes must be a negative 
energy mode. This result is known as the Krein-Moser theorem 
\cite{Moser}.   In Fig.~\ref{fig:d1} the transition that occurs at 
$\kat\approx 2.39$ and its inverse that occurs at $\kat\approx 4.91$ are both 
Krein-Moser transitions. Consequently one of the modes involved {\it 
must} be a negative energy mode, and we have shown this to be the case.  It is important to note that the 
Hamiltonian formalism is the only reliable way to determine the 
existence of negative energy modes of a system.  If one chooses an 
initial condition that only excites the drift-shear wave, the energy of the 
system will be negative.  This is essentially what is shown when  the sign of $H_L={\up_k^{(i)}}^*\hat{H}_L \up_k^{(i)}$ is obtained. 

The existence of negative energy modes is a necessary condition for 
the transition to instability, but it is not a sufficient condition. 
Upon collision, modes may merely pass through each other and remain 
on the stable axis.  The Hamiltonian is not a frame dependent 
quantity, and  it is possible to   change  the signature of a mode 
with a frame change.  For systems that possess a momentum invariant, 
the energy in the new frame is the same as the old, but with doppler 
shifted frequencies. Sometimes it is possible to remove a negative 
energy  mode by this procedure and thus obtain  a Liapunov type 
stability argument that  precludes the transition to instability 
\cite{coppi, kueny}.  In any event,  when a transition at finite 
frequency occurs, one is certain of the existence of a negative 
energy mode.

After the transition, the energy drops to zero.  Unstable modes always have zero energy, and 
thus they have no signature.  After the transition, the 
unstable gap modes have eigenvalues that occur in quartet form, $\pm 
\om_R \pm \om_I$,  and the unstable normal form is
\bqy
H_L''&=& {\sum_k}''
\Big[
\om_{Rk}  \left({P_k^{(1)}}{Q_k^{(2)}} - {P_k^{(2)}}{Q_k^{(1)}}\right)
-\om_{Ik}  \left({P_k^{(1)}}{Q_k^{(1)}} + {P_k^{(2)}}{Q_k^{(2)}}\right)
\nonumber\\
  &{\ }& \hspace{1.0 in}  + \frac1{2} \sum_{i=3,4} \si^{(i)}_k \om_k^{(i)} 
\left({P_k^{(i)}}^2 + {Q_k^{(i)}}^2\right)
  \Big]\,.
\label{UStNoFo}
\eqy
where the double prime indicates a sum over $k$-values of the gap, 
the indices 1 and 2 denote the upper Alfven and drift modes, and the 
indices 3 and 4 denote the remaining two stable modes.  In the new 
coordinates, the total Hamiltonian is given by $H_L=H_L'+H_L''$. 
Recall, in all of these sums, the sum over the $x$-components of the 
wave numbers have been suppressed.

In Fig.~\ref{fig:d2}  the unstable gap arising from the collision 
of the Alfven and drift modes occurs for values  $2.12\leq \kat\leq 3.99$,  and then 
another collision occurs  when a stable wave with positive signature collides with 
the lower drift-shear  mode at $\kat\approx 4.95$.  This lower mode changed from its positive energy value in Fig.~\ref{fig:d1}  to a negative energy mode upon traversing $N=0$.  For $\kat>4.95$,  the normal form is like (\ref{UStNoFo}),  except now the 
stable modes are the lower  Alfven wave and the mode that limits to $N_{\de}$,  and the other two modes are unstable for arbitrarily large $\kat$.

{}From Eqs.~(\ref{sig1}) and (\ref{sig}) we see that if all modes are stable they will   have positive energy if $(N_{\de}\mp N_+)(N_{\de}\mp N_-)>0$ for both signs.  When this is the case, the system is energy stable, which can be traced back the the positive definiteness of $\de^2 F$.  This kind of stability is sometimes called energy-Casimir stability (see \cite{mor98} for references to the early plasma literature where this idea was introduced).  Being energy stable means that our Hamiltonian $H_L$ can be brought into the form (\ref{StNoFo}) for all $k$, with $\si^{(i)}_k>0$ for all $i$ and $k$. 



\section{Collisionless conductivity and tearing modes}
\label{tearing}

In this section we consider another linear application. We show  that 
the Jacobi identity of the bracket `$[\ ,\ ]$', an identity essential for the Hamiltonian description of Sec.~\ref{ham_form} and its Jacobi identity for the bracket `$\{\ ,\ \}$',   can be used to  show that the azimuthal current density responds {\em locally} to the electric field on each flux surface. That is, the 
current is proportional to the parallel electric field and is 
independent of its gradient across flux surfaces. This would not be 
true in the presence of particle diffusivity or in the presence of a 
finite electron gyroradius. The locality of the conduction is 
important as it allows the linearized system of equations to be 
reduced to a system that differs from that describing resistive 
tearing modes in a sheared slab   by  merely  replacing   the 
collisional conductivity $\sigma$ by a spatially dependent, AC 
collisionless conductivity.


\subsection{Collisionless conductivity}

We consider perturbations of a plasma slab in which all equilibrium fields vary only in the $x$-direction, i.e.\  the fields variables have  the following form:
\begin{eqnarray}  
\psi&=&\psi_0(x)+\tilde{\psi}_k(x)e^{i \omega t -i k y} + c.c.\,,
\label{psif}\\
\varphi&=&-{\omega}x/k+\tilde{\varphi}_k(x)e^{i \omega t -i k 
y} + c.c.\,,  \label{phif}
\label{eqflow} \\
Z&=&\alpha_{Z}x+\tilde{Z}_k(x)e^{i \omega t -i k y} + c.c.\,, 
\label{Zf}
\\
v&=&\alpha_{v}x+\tilde{v}_k(x)e^{i \omega t -i k y} + c.c.\,, 
\label{vf}
\end{eqnarray}
where, to avoid clutter, below  we drop the subscript $k$ on the tilde variables.  The above equations are the same as those of Sec.~\ref{cancor},  except here we assume  that  $\psi_0(x)$ of  (\ref{psif}) is a polynomial of at most 
quadratic order in $x$.  This assumption  makes  it possible to  apply the final result to two cases, namely the homogeneous  equilibrium of Sec.~\ref{homeq} and the case of a magnetic equilibrium with a resonant  surface at $x=0$. The presence of an equilibrium flow in 
Eq.~(\ref{eqflow}) corresponds to choosing a reference frame moving 
with the phase velocity of the perturbation. This choice is 
convenient  because it allows us to ignore the  terms with the time 
derivatives in our calculations below.

The Jacobi identity for $[\ ,\ ]$ is
\begin{equation} \label{jac1}
[\varphi,[\psi,\xi]]+[\xi,[\varphi,\psi]]+[\psi,[\xi,\varphi]]=0\, ,
\end{equation}
where $\xi$ represents any of the field variable of 
(\ref{psif})-(\ref{vf}) and consequently it has the form 
$\xi=\xi_0(x)+\tilde{\xi}(x)\exp({i \omega t -i k y})+ c.c$.
Linearizing (\ref{jac1}) and  retaining  terms  
 of first order gives
\begin{equation} \label{linjac}
-\frac{\omega}{k}\frac{\partial}{\partial 
y}[\psi,\xi]_L+\xi_0^{\prime}\frac{\partial}{\partial 
y}[\varphi,\psi]_L+\psi_0^{\prime}\frac{\partial}{\partial 
y}[\xi,\varphi]_L=0,
\end{equation}
where $[f,g]_L$ is a short-hand notation for the linearized Poisson 
bracket between two fields, e.g.\  
$[\psi,\xi]_L=-ik(\psi_0^{\prime} \tilde{\xi} -\xi_0' 
\tilde{\psi})\exp(i \omega t -ik y)$. 

By replacing $\xi$ with $Z$ in (\ref{linjac}) and using (\ref{e2}) one obtains
\begin{equation}  \label{Z1}
-\frac{\omega}{k}\frac{\partial}{\partial y}[\psi,Z]_L+\alpha_Z 
\frac{\partial}{\partial y}[\varphi,\psi]_L+ \psi_0^{\prime} 
\frac{\partial}{\partial 
y}(d_{\beta}[J,\psi]_L-c_{\beta}[v,\psi]_L)=0,
\end{equation}
where $J=-\nabla^2 \psi$. On the other hand by replacing $\xi$ with 
$v$ in (\ref{linjac}) one gets
\begin{equation}
\frac{\partial}{\partial 
y}[v,\psi]_L=-\frac{k}{\omega}\left(\alpha_v 
\frac{\partial}{\partial y} [\varphi,\psi]_L + 
\psi_0^{\prime}\frac{\partial}{\partial y}[v,\varphi]\right)=0.
\end{equation}
The latter expression can be used to replace $\partial/\partial y 
[v,\psi]_L$ in (\ref{Z1}) and obtain
\begin{equation} \label{Z2}
-\frac{\omega}{k}\frac{\partial}{\partial y}[\psi,Z]_L+\alpha_Z 
\frac{\partial}{\partial y}[\varphi,\psi]_L+ \psi_0^{\prime} 
d_{\beta} \frac{\partial}{\partial y} [J,\psi]_L +\psi_0^{\prime} 
c_{\beta}\left( \frac{k}{\omega}\alpha_v \frac{\partial}{\partial 
y} [\varphi,\psi]_L + \frac{k}{\omega} \psi_0^{\prime} 
\frac{\partial}{\partial y}[v,\varphi]\right)=0.
\end{equation}
Making use of (\ref{e4}), (\ref{Z2}) can be reformulated as
\begin{equation} \label{Z3}
-d_{\beta}\frac{\partial}{\partial y}[\psi,Z]_L= d_{\beta} \frac{ 
(\omega \alpha_Z +\psi_0^{\prime} \alpha_v c_{\beta} 
k)\frac{\partial}{\partial y} [\varphi, \psi]_L + \omega 
\psi_0^{\prime} d_{\beta} \frac{\partial}{\partial y}[J,\psi]_L}{- 
\omega^2 +{\psi_0^{\prime}}^2 c_{\beta}^2 k^2}.
\end{equation}
Ohm's law (\ref{e1}), on the other hand, yields
\begin{equation} \label{ohm2}
-d_{\beta} \frac{\partial}{\partial 
y}[\psi,Z]_L=-\frac{\partial}{\partial y}[\varphi ,\psi]_L -d_e^2 
\frac{\partial}{\partial y} [\varphi,J].
\end{equation}
Given that in the linearized Poisson bracket the dependence on $y$ 
enters only through the exponential, the derivative with respect to 
$y$ amounts to a multiplication times $-i k$. Bearing this in mind 
and combining (\ref{Z3}) with (\ref{ohm2}), yields
\begin{equation}
\frac{d_{\beta} k \omega \alpha_Z +d_{\beta} \psi_0^{\prime} 
\alpha_v c_{\beta} k^2 -\omega^2 +{\psi_0^{\prime}}^2 c_{\beta}^2 
k^2}{-\omega^2 +{\psi_0^{\prime}}^2 c_{\beta}^2 k^2} 
[\varphi,\psi]_L=\left[ \frac{\omega k  \psi_0^{\prime} 
d_{\beta}^2}{-\omega^2 +{\psi_0^{\prime}}^2 c_{\beta}^2 k^2} \psi 
-d_e^2 \varphi,J\right]_L\,.
\end{equation}
Now if one introduces in the linearized Poisson brackets the explicit 
expressions (\ref{psif})-(\ref{vf}) for the fields, one obtains
\begin{equation}  \label{cond}
i\sigma(x) (\omega \tilde{\psi} +\psi_0^{\prime} k 
\tilde{\varphi})=-k^2 \tilde{\psi} +\tilde{\psi}''
\end{equation}
where  
\begin{equation}
\sigma(x):=\frac{
\omega \omega_*  +c_{\beta} d_{\beta} 
\psi_0^{\prime}k \omega_{KH} -\omega^2 +{\psi_0^{\prime}}^2 
c_{\beta}^2 k^2}
{i\omega(k^2 {\psi_0^{\prime}}^2 d_{\beta}^2 
+d_e^2 (-\omega^2 +{\psi_0^{\prime}}^2 c_{\beta}^2 
k^2))}\,,
\label{sigma}
\end{equation}
with $\omega_*=d_{\beta} k \alpha_Z$ and $\omega_{KH}=k \alpha_v$.
This shows that Ohm's law can be written as a proportionality 
relation between the amplitudes of the projection of the current 
density along $z$ and of the electric field along the poloidal 
magnetic field. The quantity $\sigma(x)$ then plays the role of a 
spatially dependent conductivity. 

The case $\psi_0(x)=x^2/2 L_s$  corresponds to  an 
equilibrium with scale length $L_s$ and with a resonant surface at 
$x=0$. If for this case we restrict to a thin layer around the resonant 
surface, the system comprised of    (\ref{cond}) and   the 
vorticity equation (\ref{e3}) can be approximated by
\bq
i\sigma(x)(\omega \tilde{\psi} +k_{\parallel} 
\tilde{\varphi})=\tilde{\psi}'' 
\quad{\rm and} \quad 
\omega \tilde{\phi}''(x)=-k_{\parallel} \tilde{\psi}''\,,
\label{ohminn}
\eq
where $k_{\parallel}=k x/L_s$ and  $y$-derivatives, being  negligible in the layer, have been dropped. Thus the layer equations 
for the present model take the same form as those of  MHD,
\bq
\omega\tilde{\varphi}''(x)=x\tilde{\psi}''(x) 
\quad{\rm and}\quad
\sigma(x) E_\|=\tilde{\psi}''(x)\, ,
\label{eq_linOhm}
\eq
except for the replacement of the conductivity by the  spatially 
varying AC conductivity $\sigma(x)$ of (\ref{sigma}). 
Here $E_\|=i(\omega\tilde{\psi}+x \tilde{\varphi})$.

The case  $\psi_0(x)=\alpha_{\psi}x $ corresponds to the homogeneous equilibria of Sec.~\ref{normal}. Here  $\sigma$ is constant and  Eqs.~(\ref{ohminn}) become
\bq 
i\sigma (\omega \tilde{\psi} +\alpha_{\psi} k 
\tilde{\varphi})=-k^2 \tilde{\psi} +\tilde{\psi}''
\quad{\rm and}\quad
\omega(-k^2 \tilde{\varphi} +\tilde{\varphi}'')=-\alpha_{\psi} k 
(-k^2 \tilde{\psi} +\tilde{\psi}'').
\eq
The solvability condition for this system, for solutions with 
dependence on $x$ of the form $\exp({-ik_x x})$, gives again the dispersion 
relation (\ref{dispn}).


\subsection{Collisionless tearing mode}
\label{ctm}

Now  the dispersion relation derived above is used to obtain  the 
growth rate for  collisionless tearing modes. We restrict attention to 
the case of moderate $\Delta'$ where the constant-$\tilde{\psi}$ 
approximation applies. Mirnov et al.\cite{MirnHegnPragr} have 
recently described the opposite case of large $\Delta'$ using a 
two-fluid model analogous to that of  \cite{Fit04}.

In the constant-$\tilde{\psi}$ approximation,  the dispersion 
relation for the tearing mode follows from the matching condition for 
the magnetic perturbation,
\bq
\Delta'=\frac{1}{\tilde{\psi}}\int_{-\infty}^{\infty}dx\,\tilde{J}\,.
\label{cch}
\eq
At the resonant surface, $k_\|=0$, the conductivity is very high due 
to the high electron mobility, 
$i\omega\sigma(0)=(1-\omega_*/\omega)/d_e^2$. Away from the resonant 
surface, however, the conductivity decreases rapidly due to the 
shielding of the electric field by the electron motion along the 
magnetic field. The shielding is described by the $k_\|d_\beta$ term 
in the denominator. The region of high conductivity is called the 
current channel and for moderate tearing parameter $\Delta'$, it 
contains most of the current in the reconnection layer. In the 
current channel, the conductivity may be approximated by
\[\sigma(x)\simeq  - d_e^{-2}\frac{1-\omega_*/\omega}{
    i\omega\left[ \frac{k_\|^2d_\beta^2}{\omega^2d_e^2}-1\right]},\]
Substituting this in the matching integral and evaluating the 
integral gives 
\[
\Delta'=-i\pi(\omega-\omega_*)\,\frac{L_s}{kd_\beta d_e}\,,
\]
whence we  obtain the dispersion relation 
\bq
\omega =\omega_*+i\frac{\Delta'kd_\beta}{\pi L_s}{d_e}.
\label{ddr}
\eq
In the limit $\beta\ll 1$ (\ref{ddr})  agrees with the kinetic 
result of  \cite{DrakeLee_lin}, aside from a factor 
of $2/\sqrt{\pi}\simeq 1.13$ in the growth rate, and it agrees with the fluid result of \cite{GOP02} in the low $\beta$ limit.



\section{Saturation of  the collisionless tearing mode}
\label{saturation}

For our  last application of the Hamiltonian formalism we use a  Casimir invariant  to  find  the nonlinear saturated state of the collisionless tearing mode following an approach similar to that in \cite{Wael89},  except that here we make use of the constant-$\psi$ approximation to simplify the analysis.   For simplicity we consider the cold plasma limit  where $\db=\cb=0$.   Inspection of (\ref{e1}) reveals that  the Casimir $C_2$ of (\ref{C2}) survives but with $D$ replaced by $\psi_e$,   Eq.~(\ref{e3}) is unaltered, and Eqs.~(\ref{e2}) and (\ref{e4}) ensure that if initially $Z,v\equiv 0$, then they will remain so.  

We use the invariance of $C_2$  to describe what becomes of an unstable un-reconnected state, $\psi^{(0)}={B_0}x^2/2{L_s}$, such as that described in Sec.~\ref{ctm}, as it evolves into a final approximate equilibrium state  $\psi^{(\infty)}= {B_0}{x^2}/{2L_s}+\bar{\psi}\cos y$ that represents  a magnetic island of half-width  $w=(4L_s\bar{\psi}/B_0)^{1/2}$.  For convenience, throughout this section we use dimensionless units where lengths are scaled with $w$ and the flux with $B_0/L_s$. In these units, for example, $\psi^{(\infty)}$ becomes $\psi^{(\infty)}=  {x^2}/{2}+ \cos y/4$.

Choosing $\mathcal{K}(\psi_e)=\delta(\psi_e-\hat{\psi}_e)/2\pi$ singles out the surface of constant $\hat{\psi}_e$, yielding 
\bq
C_2(\hat{\psi}_e)=\frac{1}{2\pi}\oint \frac{dy}{\partial_x\psi_e}\,,
\label{c2d}
\eq
where $\partial_x\psi_e(x,y,t)$ is to be evaluated at $x=\psi_e^{-1}(\hat{\psi}_e,y,t)$, and this can be done  at any time.  From  $\psi^{(0)}$ we obtain $\psi_e^{(0)}= x^2/2 -\eta^2$, where $\eta^2:=d_e^2/w^2$;  whence for the initial state $C_2(\hat{\psi}_e)=  {1}/{\sqrt{2(\hat{\psi}_e+\eta^2)}}$. 
  
Because the final state is an equilibrium state, $\psi_e^{(\infty)}$ is a function of ${\psi^{(\infty)}}$ according to ${\psi_e^{(\infty)}}= {\psi^{(\infty)}} +\eta^2 I({\psi^{(\infty)}})$, where $I$ is the final current profile  of the saturated island.  Thus (\ref{c2d}) becomes 
\bq
C_2(\hat{\psi}_e)= 
                \left(\frac{d\psi_e^{(\infty)}}{d\psi^{(\infty)}}\right)^{-1}\!\!\frac1{2\pi}\oint 
\frac{dy}{ \partial_x\psi^{(\infty)}}\,,
\label{c2f}
\eq
where  $\partial_x\psi^{(\infty)}(x,y)$ is to be evaluated at $x=(\psi_e^{(\infty)})^{-1}(\hat{\psi}_e,y)$ and  ${d\psi_e^{(\infty)}}/{d\psi^{(\infty)}}$, being expressible as a function of $\hat{\psi}_e$ alone, can be pulled outside the integral.  The  final current profile   follows by setting the initial and
final values  of $C_2(\hat{\psi}_e)$ equal at each $\hat{\psi}_e$, yielding 
\bq
\frac{1}{\sqrt{2(\hat{\psi}_e+\eta^2)}}\frac{d\hat{\psi}_e}{d\psi^{(\infty)}}=\frac1{2\pi}\oint 
\frac{dy}{\partial_x\psi^{(\infty)}}\,. 
\label{Ide}
\eq
Although $\psi^{(\infty)}$ is  unknown, for small $\Delta'$,  $\p_x\psi^{(\infty)}\approx x\approx \sqrt{(4\psi^{(\infty)} - \cos y)/2}$.  Integrating   Eq.~(\ref{Ide}) yields,
\bq
\hat{\psi}_e=\iota^2(\psi^{(\infty)})-\eta^2\,.
\eq
where 
\[\iota(\psi^{(\infty)}) = \frac{1}{2\pi}\int_0^{\pi} dy \, \sqrt{4\psi^{(\infty)} -\cos y}.\]
and where we have used the fact that $\lim_{\psi\rightarrow\infty} (\hat{\psi}_e(\psi)-\psi)=0$ to set the integration constant to zero. 
 
 Using $\psi_e^{(\infty)}= \psi^{(\infty)}+\eta^2 I(\psi^{(\infty)})$, and  dropping  the `hat' and the label $\infty$, gives an equation for the current profile,
 \bq
 I(\psi) = -1 + \eta^{-2}\left[-\psi+\iota^2(\psi)\right]
 \label{answ}
 \eq
The function  $\iota(\psi)$  is easily evaluated in terms of elliptic integrals:
\begin{equation*}
\iota(\psi)=\left\{\begin{array}{ll}
\frac{2}{\pi}\sqrt{2\psi+1/2}\,E(1/(2\psi+1/2)), & \mbox{ for } \psi>1/4;\\
\sqrt{2}\left(E(2\psi+1/2) + (2\psi-1/2)K(2\psi+1/2)\right), & \mbox{ for } -1/4<\psi<1/4\,, 
\end{array}\right.
\end{equation*}
where $K$ and $E$ are the complete elliptic integral of the first and second  kind. 

\begin{figure}[htbp]
  \begin{center}
    \includegraphics[width=.5\linewidth]{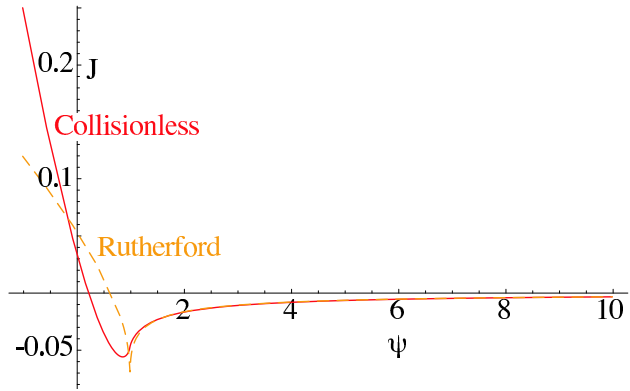}
  \end{center}
  \caption{Comparison of the current density profiles, one  determined 
by parallel electron momentum conservation and the other 
by Ohmic diffusion (dashed line)}
  \label{fig:J}
\end{figure}

Figure \ref{fig:J} shows a comparison of the above current profile with  that of  Rutherford \cite{ruth73}
for   resistive diffusion. Observe, the  profiles are qualitatively similar.
Substituting the current profile in the matching relation of (\ref{cch}) yields the 
saturation amplitude $w=\Delta'd_e^2/G$, where
\[
G=8\int_{\psi_{\rm min}}^\infty d\psi\,I(\psi) \oint  \frac{dy}{2\pi}\,\frac{\cos y}{\partial_x\psi}=0.19
\]
This is close to the value $G=0.205$ obtained by Drake and Lee 
\cite{DrakeLee_nonlin} using a kinetic model.  

 
 
\section{Summary and conclusions}
\label{conclusions}

In the early sections of this paper we presented the  noncanonical Hamiltonian formulation of the 
four-field model of  \cite{Fit04}, and showed that the associated  
Lie-Poisson bracket has four new independent families of 
Casimir invariants.  These  invariants led us to the discovery of variables in which the Poisson bracket has the simple form (\ref{clean_bkt}), and in which the system can be written in the compact form of (\ref{eq_Ddot})-(\ref{eq_Tpm}). 

In Sec.~\ref{equilibria} we used the Hamiltonian formulation to obtain a variational principle that gives  a  set of coupled differential equations that 
generalize the Grad-Shafranov equilibrium equation. In the limit of 
vanishing electron mass ($d_e\rightarrow 0$) the equilibrium 
equations reduce to previously known results. This limit provides 
some insight into the relationship of the Casimirs to the more 
familiar conserved quantities of conventional low-$\beta$ drift 
models.  We have presented two solutions of the equilibrium 
equations, the first describing dipole-like equilibria and the second describing  homogeneous equilibria that support drift-acoustic and Alfv\'en modes.   

In Sec.~\ref{normal} we investigated the linear dispersion relation for homogeneous  equilibria and described the map to the appropriate Hamiltonian forms.  We also presented thresholds for spectral and energy stability.  In Sec.~\ref{stabsignf} we described a method for determining the energy signature of a mode.   This method is of general utility and can be applied to all valid models, provided one understands their Hamiltonian structure.  In obtaining reduced fluid models,   there can be ambiguity about the energy for the full dynamics, and linear theory alone cannot be used to uniquely determine the correct energy of the linear dynamics.  The only reliable way to determine the energy is from a Hamiltonian or action principle formulation, where the energy for the linear dynamics is obtained by expansion of a Hamiltonian associated with time translation symmetry.  
 
In Secs.~\ref{tearing} and \ref{saturation} we demonstrated the usefulness of the Hamiltonian 
formulation for the analysis of the linear  collisionless tearing mode and its nonlinear
saturation. In the case of linear stability, the Jacobi identity allows the reduction of the system to 
a form analogous to that of  MHD but where the conductivity is replaced by a spatially 
varying AC conductivity. Applications of the formalism left 
for future work include  the study of the stability of the saturated states 
against secondary modes and an investigation of saturation in a more general dynamical context  
using additional Casimirs.

Another area for future work concerns the families associated with  the invariants 
$T_{\pm}$, which  generalize a pair of Casimirs $G_{\pm}$ found  for a low-$\beta$ two-field model derived in \cite{Sch94}.  A natural question is whether the invariants $T_{\pm}$ play a role 
analogous to the one played by $G_{\pm}$ in the two-field limit in 
determining the alignment of current density and vorticity along the 
separatrices of the magnetic field during the nonlinear evolution of 
the system \cite{Caf98}. The  present model makes possible an  investigation 
of this question  along  the lines carried out in \cite{Gra01}.

\section*{Acknowledgments}

Two of us (ET and DR) were partly supported by the Euratom Communities under the
contract of Association between EURATOM/ENEA. The views and opinions
expressed herein do not necessarily reflect those of the European
Commission.  These two would also like to thank F.\   Pegoraro for   drawing  their  
attention to this problem and   to thank  T.\  J.\  Schep for useful discussions.  The other two of us  (PJM and FLW) were 
supported  by the US Department of Energy Contract No.~DE-FG03-96ER-54346.  One of us (PJM)  would like to thank  W.\   Horton for useful discussions.

\bibliographystyle{unsrt}

\bibliography{7.6wrk}

\begin{thebibliography}{10}

\bibitem{RogrDentDrak}
B.~N. Rogers, R.~E. Denton, J.~F. Drake, and M.~A. Shay.
\newblock Role of dispersive waves in collisionless magnetic reconnection.
\newblock {\em Phys. Rev. Lett.}, 87:195004, 2001.

\bibitem{Pri00}
E.~R. Priest and T.~G. Forbes.
\newblock {\em Magnetic Reconnection}.
\newblock Cambridge University Press, 2000.

\bibitem{Bis00}
D.~Biskamp.
\newblock {\em Magnetic Reconnection in Plasmas}.
\newblock Cambridge University Press, 2000.

\bibitem{Sch94}
T.~J. Schep, F.~Pegoraro, and B.~N. Kuvshinov.
\newblock Generalized two-fluid theory of nonlinear magnetic structures.
\newblock {\em Phys. Plasmas}, 1:2843--2851, 1994.

\bibitem{DastMaWeilnd}
S.~Dastgeeer, S.~Mahajan, and J.~Weiland.
\newblock Zonal flows and transport in ion temperature gradient turbulence.
\newblock {\em Phys. Plasmas}, 9:4911--4916, 2002.

\bibitem{KromKolesh}
J.~A. Krommes and R.~A. Kolesnikov.
\newblock Hamiltonian description of convective-cell generation.
\newblock {\em Phys. Plasmas}, 11:L29--L32, 2004.

\bibitem{KolesKromPRL}
R.~A. Kolesnikov and J.~A. Krommes.
\newblock Transition to collisionless ion-temperature-gradient-driven plasma
  turbulence: A dynamical systems approach.
\newblock {\em Phys. Rev. Lett.}, 94:235002, 2005.

\bibitem{KolesKromPoP}
R.~A. Kolesnikov and J.~A. Krommes.
\newblock Bifurcation theory of the transition to collisionless
  ion-temperature-gradient-driven plasma turbulence.
\newblock {\em Phys. Plasmas}, 12:122302, 2005.

\bibitem{Xu_L-H}
X.Q. Xu, R.~H. Cohen, T.~D. Rognlien, and J.~R. Myra.
\newblock Low-to-high confinement transition simulations in divertor geometry.
\newblock {\em Phys. Plasmas}, 7:1951--1958, 2000.

\bibitem{RogrDrakZeilr}
B.~N. Rogers, J.~F. Drake, and A.~Zeiler.
\newblock Phase space of tokamak edge turbulence, the l-h transition, and the
  formation of the edge pedestal.
\newblock {\em Phys. Rev. Lett.}, 81:4396--4399, 1998.

\bibitem{GuzMaYo}
P.~N. Guzdar, S.~M. Mahajan, and Z.~Yoshida.
\newblock A theory for the pressure pedestal in high ({H}) mode tokamak
  discharges.
\newblock {\em Phys. Plasmas}, 12:032502, 2005.

\bibitem{ScottRvw08}
Bruce~D Scott.
\newblock Tokamak edge turbulence: background theory and computation.
\newblock {\em Plasma Phys. Control. Fusion}, 49:S25--S41, 2007.

\bibitem{ScottGEM06}
Bruce~D. Scott.
\newblock Free-energy conservation in local gyrofluid models.
\newblock {\em Phys. Plasmas}, 12:102307, 2005.

\bibitem{mor82}
P.~J. Morrison.
\newblock Poisson brackets for fluids and plasmas.
\newblock In M.~Tabor and Y.~Treve, editors, {\em Mathematical Methods in
  Hydrodynamics and Integrability in Dynamical Systems}, volume~88 of {\em
  American Institute of Physics Conference Proceedings}, pages 13--45. American
  Institute of Physics, 1982.

\bibitem{mor98}
P.~J. Morrison.
\newblock Hamiltonian description of the ideal fluid.
\newblock {\em Rev. Mod. Phys.}, 70:467--521, 1998.

\bibitem{mor05}
P.~J. Morrison.
\newblock Hamiltonian and action principle formulations of plasma physics.
\newblock {\em Phys. Plasmas}, 12:058102--1--058102--13, 2005.

\bibitem{MG80}
P.~J. Morrison and J.~M. Greene.
\newblock Noncanonical {H}amiltonian density formulation of hydrodynamics and
  ideal magnetohydrodynamics.
\newblock {\em Phys. Rev. Lett.}, 45:790--793, 1980.
\newblock Erratum: \textbf{48}, 569 (1982).

\bibitem{MH84}
P.~J. Morrison and R.~D. Hazeltine.
\newblock Hamiltonian formulation of reduced magnetohydrodynamics.
\newblock {\em Phys. Fluids}, 27:886--897, 1984.

\bibitem{mar_mor84}
J.~E. Marsden and P.~J. Morrison.
\newblock Noncanonical {H}amiltonian field theory and reduced {M}{H}{D}.
\newblock {\em Contemp. Math.}, 28:133--13, 1984.

\bibitem{Haz87}
R.~D. Hazeltine, C.~T. Hsu, and P.~J. Morrison.
\newblock Hamiltonian four-field model for nonlinear tokamak dynamics.
\newblock {\em Phys. Fluids}, 30:3204--3211, 1987.

\bibitem{Kuv94}
B.~N. Kuvshinov, F.~Pegoraro, and T.~J. Schep.
\newblock Hamiltonian formulation of low-frequency, nonlinear plasma dynamics.
\newblock {\em Phys. Lett. A}, 191:296--300, 1994.

\bibitem{Gra01}
D.~Grasso, F.~Califano, F.~Pegoraro, and F.~Porcelli.
\newblock Phase mixing and saturation in {H}amiltonian reconnection.
\newblock {\em Phys. Rev. Lett.}, 86:5051--5054, 1994.

\bibitem{Gra99}
D.~Grasso, F.~Califano, F.~Pegoraro, and F.~Porcelli.
\newblock Hamiltonian magnetic reconnection.
\newblock {\em Plasma Phys. Control. Fusion}, 41:1497--, 1999.

\bibitem{WaelMorHor}
F.~L. Waelbroeck, P.~J. Morrison, and W.~Horton.
\newblock Hamiltonian formulation and coherent structures in electrostatic
  turbulence.
\newblock {\em Plasma Phys. Control. Fusion}, 46:1331--1350, 2004.

\bibitem{HM}
R.~D. Hazeltine and J.~D. Meiss.
\newblock Shear-alfven dynamics of toroidally confined plasmas.
\newblock {\em Phys. Repts.}, 121:1--164, 1985.

\bibitem{Porcelrvw}
F.~Porcelli, D.~Borgogno, F.~Califano, D.~Grasso, M.~Ottaviani, and
  F.~Pegoraro.
\newblock Recent advances in collisionless magnetic reconnection.
\newblock {\em Plasma Phys. Control. Fusion}, 44:B389--B405, 2002.

\bibitem{KuvLakPegSchep}
B.~N. Kuvshinov, V.~P. Lakhin, F.~Pegoraro, and T.~J. Schep.
\newblock Hamiltonian vortices and reconnection in a magnetized plasma.
\newblock {\em J. Plasma Phys.}, 59:727--736, 1998.

\bibitem{AYA}
A.~Y. Aydemir.
\newblock Nonlinear studies of m=1 modes in high-temperature plasmas.
\newblock {\em Phys. Fluids B}, 4:3469--3472, 1992.

\bibitem{KDW}
Robert~G. Kleva, J.~F. Drake, and F.~L. Waelbroeck.
\newblock Fast reconnection in high temperature plasmas.
\newblock {\em Phys. Plasmas}, 2:23--34, 1995.

\bibitem{Por93}
M.~Ottaviani and F.~Porcelli.
\newblock Nonlinear collisionless magnetic reconnection.
\newblock {\em Phys. Rev. Lett.}, 71:3802--3805, 1993.

\bibitem{Caf98}
E.~Cafaro, D.~Grasso, F.~Pegoraro, F.~Porcelli, and A.~Saluzzi.
\newblock Invariants and geometric structures in nonlinear {H}amiltonian
  magnetic reconnection.
\newblock {\em Phys. Rev. Lett.}, 80:4430--4433, 1998.

\bibitem{DelsCaliPeg}
D.~Del Sarto, F.~Califano, and F.~Pegoraro.
\newblock Electron parallel compressibility in the nonlinear development of
  two-dimensional collisionless magnetohydrodynamic reconnection.
\newblock {\em Mod. Phys. Lett. B}, 20:931--961, 2007.

\bibitem{VeksBian}
G.~Vekstein and N.~H. Bian.
\newblock Hall assisted forced magnetic reconnection.
\newblock {\em Phys. Plasmas}, 13:122105, 2006.

\bibitem{BianVeks}
N.~Bian and G.~Vekstein.
\newblock On the two-fluid modification of the resistive tearing instability.
\newblock {\em Phys. Plasmas}, 14:072107, 2007.

\bibitem{RogrKobaRic}
B.~N. Rogers, S.~Kobayashi, P.~Ricci, W.~Dorland, J.~Drake, and T.~Tatsuno.
\newblock Gyrokinetic simulations of collisionless magnetic reconnection.
\newblock {\em Phys. Plasmas}, 14:092110, 2007.

\bibitem{Fit04}
R.~Fitzpatrick and F.~Porcelli.
\newblock Collisionless magnetic reconnection with arbitrary guide field.
\newblock {\em Phys. Plasmas}, 11:4713--4718, 2004.
\newblock Erratum: \textbf{14}, 049902 (2007).

\bibitem{T07}
E.~Tassi, P.~J. Morrison, and D.~Grasso.
\newblock Hamiltonian structure of a collisionless reconnection model valid for
  high and low $\beta$ plasmas.
\newblock In G.~Bertin, R.~Pozzoli, M.~Rome, and K.~R. Sreenivasan, editors,
  {\em Collective phenomena in macroscopic systems}, pages 197--206. World
  Scientific, 2007.

\bibitem{ruth73}
P.~H. Rutherford.
\newblock Nonlinear growth of the tearing mode.
\newblock {\em Phys. Fluids}, 16:1903--1908, 1973.

\bibitem{Thi00}
J.~L. Thiffeault and P.~J. Morrison.
\newblock Classification and {C}asimir invariants of {L}ie-{P}oisson brackets.
\newblock {\em Physica D}, 136:205--244, 2000.

\bibitem{YoshiMaha}
Z.~Yoshida and S.~M. Mahajan.
\newblock Variational principles and self-organization in two-fluid plasmas.
\newblock {\em Phys. Rev. Lett.}, 88:095001, 2002.

\bibitem{YoshiMahaOhsa}
Z.~Yoshida, S.~M. Mahajan, and S.~Ohsaki.
\newblock Scale hierarchy created in plasma flow.
\newblock {\em Phys. Plasmas}, 11:3660--3664, 2004.

\bibitem{Goedb04}
J.~P. Goedbloed.
\newblock Variational principles for stationary one- and two-fluid equilibria
  of axisymmetric laboratory and astrophysical plasmas.
\newblock {\em Phys. Plasmas}, 11:L81--L84, 2004.

\bibitem{HiroYoshHamri}
M.~Hirota, Z.~Yoshida, and E.~Hameiri.
\newblock Variational principle for linear stability of flowing plasmas in hall
  magnetohydrodynamics.
\newblock {\em Phys. Plasmas}, 13:022107, 2006.

\bibitem{HamriTora}
Eliezer Hameiri and R.~Torasso.
\newblock Linear stability of static equilibrium states in the
  hall-magnetohydrodynamics model.
\newblock {\em Phys. Plasmas}, 11:4934--4945, 2004.

\bibitem{ItoRamoNaka}
Atsushi Ito, Jes\'{u}s~J. Ramos, and Noriyoshi Nakajima.
\newblock Ellipticity of axisymmetric equilibria with flow and pressure
  anisotropy in single-fluid and hall magnetohydrodynamics.
\newblock {\em Phys. Plasmas}, 14:062502, 2007.

\bibitem{SteinIshi06a}
Loren~C. Steinhauer and Akio Ishida.
\newblock Nearby-fluids equilibria. {I}. {F}ormalism and transition to
  single-fluid magnetohydrodynamics.
\newblock {\em Phys. Plasmas}, 13:052513, 2006.

\bibitem{SteinGuo06b}
L.~C. Steinhauer and H.~Y. Guo.
\newblock Nearby-fluids equilibria. {II}. {Z}onal flows in a high-beta,
  self-organized plasma experiment.
\newblock {\em Phys. Plasmas}, 13:052514, 2006.

\bibitem{Gardner_kdv}
C.~S. Gardner.
\newblock {K}orteweg-de {V}ries equation and generalizations. {IV}. {T}he
  {K}orteweg-de {V}ries equations as a {H}amiltonian system.
\newblock {\em J. Math. Phys.}, 12:1548--1551, 1971.

\bibitem{Will}
J.~Williamson.
\newblock On an algebraic problem concerning the normal forms of linear
  dynamical systems.
\newblock {\em Am. J. Math.}, 58:141--163, 1936.

\bibitem{Meyer}
A.~J. Laub and K.~Meyer.
\newblock Canonical forms for symplectic and {H}amiltonian matrices.
\newblock {\em Celestial Mech.}, 9:213--238, 1974.

\bibitem{MK90}
P.~J. Morrison and M.~Kotschenreuther.
\newblock The free energy principle, negative energy modes, and stability.
\newblock In V.~G. Baryakhtar, V.~M. Chernousenko, N.~S. Erokhin, A.~B.
  Sitenko, and V.~E. Zakharov, editors, {\em Nonlinear World: IV International
  Workshop on Nonlinear and Turbulent Processes in Physics}. World Scientific,
  1990.

\bibitem{Moser}
J.~Moser.
\newblock New aspects in the theory of stability of {H}amiltonian systems.
\newblock {\em Comm. Pure Appl. Math.}, 11:81--114, 1958.

\bibitem{coppi}
B.~Coppi, M.~N. Rosenbluth, and R.~N. Sudan.
\newblock Non-linear interactions of positive and negative energy modes.
\newblock {\em Ann. Phys.}, 55:248--270, 1969.

\bibitem{kueny}
C.~S. Kueny and P.~J. Morrison.
\newblock Nonlinear instability and chaos in plasma wave-wave interactions.
  {I}. {I}ntroduction.
\newblock {\em Phys. Plasmas}, 2:1926--1940, 1995.

\bibitem{MirnHegnPragr}
V.~V. Mirnov, C.~C. Hegna, and S.~C. Prager.
\newblock Two-fluid tearing instability in force-free magnetic configuration.
\newblock {\em Phys. Plasmas}, 11:4468--4482, 2004.

\bibitem{DrakeLee_lin}
J.~F. Drake and Y.~C. Lee.
\newblock Kinetic theory of tearing instabilities.
\newblock {\em Phys. Fluids}, 20:1341--1353, 1977.

\bibitem{GOP02}
D.~Grasso, M.~Ottaviani, and F.~Porcelli.
\newblock Growth and stabilization of drift-tearing modes in weakly collisional
  plasmas.
\newblock {\em Nuc. Fusion}, 42:1067--, 2002.

\bibitem{Wael89}
F.~L. Waelbroeck.
\newblock Current sheets and nonlinear growth of the m=1 kink-tearing mode.
\newblock {\em Phys. Fluids B}, 1:2372--2380, 1989.

\bibitem{DrakeLee_nonlin}
J.~F. Drake and Y.~C. Lee.
\newblock Nonlinear evolution of collisionless and semicollisional tearing
  modes.
\newblock {\em Phys. Rev. Lett.}, 39:453--456, 1977.

\end{thebibliography}

\end{document}